\documentclass[%
reprint,
superscriptaddress,
grouptaddress,
amsmath,amssymb,
aps,
prd,
floatfix,
]{revtex4-2}
\usepackage{graphicx}% Include figure files
\usepackage{dcolumn}% Align table columns on decimal point
\usepackage{bm}%
\usepackage{physics}%
\usepackage{subcaption}%
\usepackage{cancel}%
\usepackage{xcolor}%
\usepackage[normalem]{ulem}%
\usepackage{tikz}%
\usepackage{tikz-3dplot}%
\usepackage[acronym,nomain,nolong,nosuper]{glossaries}%
\usepackage[dvipsnames,table]{xcolor}
\usepackage{hyperref}
\hypersetup{
    colorlinks=true,
    linktoc=all,
    linkcolor=RoyalBlue,
    citecolor=Rhodamine,
    urlcolor=Rhodamine,
}
\usepackage{orcidlink}
\usepackage{cleveref}
\crefname{equation}{Eq.}{Eqs.}
\crefname{figure}{Fig.}{Figs.}
\crefname{table}{Table}{Tables}
\newacronym{lisa}{LISA}{Laser Interferometer Space Antenna}
\newacronym{athena}{Athena}{Advanced Telescope for High ENergy Astrophysics}
\newacronym{gw}{GW}{gravitational wave}
\newacronym{tdi}{TDI}{time-delay interferometry}
\newacronym[longplural={Galactic binaries}]{gb}{GB}{Galactic Binary}
\newacronym[longplural={verification Galactic binaries}]{vgb}{VGB}{verification Galactic binary}
\newacronym[longplural={massive black hole binaries}]{mbhb}{MBHB}{massive black hole binary}
\newacronym{emri}{EMRI}{extreme mass ratio inspiral}
\newacronym{sgwb}{SGWB}{stochastic gravitational wave background}
\newacronym{snr}{SNR}{signal-to-noise ratio}
\newacronym{tm}{TM}{test mass}
\newacronym{oms}{OMS}{optical metrology system}
\newacronym[longplural={movable optical sub-assemblies}]{mosa}{MOSA}{movable optical sub-assembly}
\newacronym[longplural={power spectral densities}]{psd}{PSD}{power spectral density}
\newacronym[longplural={cross spectral densities}]{csd}{CSD}{cross spectral density}
\newacronym[longplural={amplitude spectral densities}]{asd}{ASD}{amplitude spectral density}
\newacronym{gpu}{GPU}{graphics processing unit}
\newacronym{ldc}{LDC}{LISA Data Challenge}
\newacronym{pta}{PTA}{Pulsar Timing Array}
\newacronym{icrs}{ICRS}{International Celestial Reference System}
\newacronym{uksa}{UKSA}{UK Space Agency}
\newacronym{llm}{LLM}{Large Language Model}
\begin{document}

%--------------------------------------
% Abstract
%--------------------------------------
\begin{abstract}
The \acrfull{lisa} is a future space-based \acrfull{gw} observatory expected to fly in the next decade.
Central to \acrshort{lisa} data analysis is \acrfull{tdi}, a pre-processing technique which reduces otherwise overwhelming laser frequency noise.
In previous work, we explored an alternative combination, coronagraphic \acrshort{tdi}, a sky-dependent null-stream channel which has the property of canceling the \acrshort{gw} signal of a point-like source when pointed at its sky position.
Because this property relies only on the geometric response of \acrshort{lisa}, coronagraphic \acrshort{tdi} is a candidate for model-agnostic sky localization.
In this work, we address three challenges which arise when analyzing realistic \acrshort{lisa} data: \textit{(i)} spacecraft motion; \textit{(ii)} signal superposition; \textit{(iii)} instrumental noise.
We derive a second-generation coronagraphic \acrshort{tdi} variable and its analytic noise model, and show that it reduces laser frequency noise in the presence of spacecraft motion while still recovering the position of the source.
Using pairs of \acrfullpl{gb} as test cases, we find that sources can be localized independently provided they do not overlap in frequency, even when they share the same sky position, whereas sources overlapping in frequency can only be localized if one dominates by a factor of ${\sim 100}$ in amplitude.
This is confirmed by the analysis of a simulation containing a \acrfull{mbhb} and 49 verification \acrshortpl{gb}.
Finally, we introduce a statistical test based on the analytic noise model, which provides well-calibrated confidence regions for the sky position of the source.
Applying it to \acrshortpl{gb} in the presence of instrumental noise, we find that the geometric response alone only weakly constrains the sky position: even at signal-to-noise ratios of order $10^4$, 90\,\% confidence regions span ${\sim 10^3}$\,deg$^2$.
These are fundamental steps towards concrete applications of coronagraphic TDI to LISA data, such as sky localization and glitch discrimination, which we discuss in this paper.
\end{abstract}

%--------------------------------------
%--------------------------------------
\title{Coronagraphic time-delay interferometry:\\
fundamental steps towards concrete applications}%
%--------------------------------------
\author{Raissa Costa Barroso\,\orcidlink{0009-0008-2703-771}}%
\email{raissa.costa-barroso@port.ac.uk}%
\affiliation{%
Institute of Cosmology and Gravitation, University of Portsmouth, Portsmouth, PO1 3FX, UK}%
\affiliation{%
Université de Caen Normandie, ENSICAEN, CNRS/IN2P3, LPC Caen UMR6534, F-14000 Caen, France}%
%--------------------------------------
\author{Yves Lemière\,\orcidlink{0000-0001-7943-1521}}%
\affiliation{%
Université de Caen Normandie, ENSICAEN, CNRS/IN2P3, LPC Caen UMR6534, F-14000 Caen, France}%
%--------------------------------------
\author{Ian Harry\,\orcidlink{0009-0008-2703-771}}%
\affiliation{%
Institute of Cosmology and Gravitation, University of Portsmouth, Portsmouth, PO1 3FX, UK}%
%--------------------------------------
\date{\today}
%--------------------------------------

\maketitle

%--------------------------------------
% Introduction
%--------------------------------------
\section{Introduction}

% LISA
The \gls{lisa} is a future space-based \gls{gw} observatory which is expected to be flying by the next decade.
\gls{lisa} consists of three spacecraft which form a nearly equilateral triangle, the \gls{lisa} constellation, following a heliocentric orbit, either leading or trailing the Earth.
With ${2.5 \times 10^6}$\,km long arms, \gls{lisa} will be able to detect \glspl{gw} in the {mHz} band, thus opening a new window in the \gls{gw} spectrum.
If \gls{lisa} offers the opportunity to observe an unprecedented wealth of physics, it also raises complex data-analysis challenges (see~\cite{colpi_lisa_2024} for a review).

% Coronagraphic TDI
Central to \gls{lisa} data analysis is \gls{tdi}, a pre-processing technique which reduces otherwise overwhelming laser frequency noise~\cite{colpi_lisa_2024}. 
It relies on combining and delaying the interferometric measurements from \gls{lisa} such that the laser noise contained in each measurement cancels out (see~\cite{tinto_time-delay_2020} for a review).
Many combinations exist~\cite{armstrong_time-delay_1999,vallisneri_geometric_2005}, but data analysis is usually performed on Michelson variables ${(X, Y, Z)}$ or quasi-orthogonal variables ${(A,E,T)}$~\cite{PhysRevD.66.122002}.
An alternative \gls{tdi} variable, coronagraphic \gls{tdi}, stands out by its \textit{sky-dependence}. 
Unlike other \gls{tdi} combinations, it depends on a set of parameters ${(f, \delta, \alpha)}$, where $f$ is the frequency, $\delta$ the declination and $\alpha$ the right ascension, and possesses the singular property of canceling \gls{gw} signal when ${(f,\delta, \alpha) \rightarrow (f_\star, \delta_\star, \alpha_\star)}$ for a point-like \gls{gw} source at position ${(\delta_\star, \alpha_\star)}$ emitting in frequency $f_\star$~\cite{gursel_near_1989,nayak_optimizing_2003,tinto_lisa_2004,vinet_physical_2007,costa_barroso_coronagraphic_2025}.
We refer to this property as the \textit{coronagraphic property} because, conceptually, it turns this \gls{tdi} combination into a coronagraph which blocks signal when pointing at a \gls{gw} source.

% Motivation
This property motivates the study of coronagraphic \gls{tdi} as a model-agnostic sky-localization method for \gls{lisa}.
It relies only on the coherent geometric response of the instrument embedded in \gls{tdi} and therefore holds regardless of the nature of the source.
This means that a coronagraphic-\gls{tdi}-based analysis bypasses the need for any waveform template which is, in principle, an advantage.
In the landscape of \gls{lisa} data analysis, such a method could find applications in \textit{low-latency} analyses, where sky location is a crucial parameter to be communicated to facilities observing in the electromagnetic domain.
Other possibilities include using coronagraphic \gls{tdi}'s sky location as a prior to parameter estimation or relying on it to identify instrumental glitches.
Such applications are driven by investigations of analogous \textit{null-stream channels} carried out in the context of \glspl{pta} and ground-based detectors.
A similar sky-localization method has been proposed in~\cite{zhu_detection_2016,10.1093/mnras/sty892} for \glspl{pta}.
For ground-based detectors, using a null-stream channel as a \gls{gw} discriminator against instrumental glitches is the main application discussed in the literature~\cite{chatterji_coherent_2006,wen_coherent_2005,ajith_null-stream_2006,harry_targeted_2011}, although tests of general relativity have also been proposed in~\cite{chatziioannou_model-independent_2012,hagihara_probing_2018}.
In the context of \gls{lisa}, however, this technique has not reached a similar level of maturity. 

% Previous paper
Previous theoretical work~\cite{nayak_optimizing_2003,tinto_lisa_2004,vinet_physical_2007} derived such a null-stream channel for \gls{lisa} but did not assess its applicability to realistic scenarios.
Our previous paper~\cite{costa_barroso_coronagraphic_2025} was a first step towards bridging this gap.
By considering a modern \gls{lisa} configuration, we characterized the method on typical \gls{lisa} sources, namely \glspl{gb} and \glspl{mbhb}.
We have also derived an analytic noise model for the coronagraphic combination which we validated on state-of-the-art simulations of the instrument.
Finally, we investigated the impact the motion of \gls{lisa} has on such sky-localization technique.

% This paper
Although the results in~\cite{costa_barroso_coronagraphic_2025} were encouraging, three crucial aspects of realistic data were not addressed:
\begin{enumerate}
    \item \textit{Second-generation \gls{tdi}:} In its original formulation, coronagraphic \gls{tdi} was derived for first-generation Sagnac variables.
    However, first-generation combinations are known to fail to suppress laser frequency noise beyond the static constellation approximation~\cite{tinto_time_2004}.
    In practice, only second-generation variables, which take into account the relative motion between spacecraft, are able to reduce laser frequency noise to acceptable levels~\cite{PhysRevD.105.062006}.
    The same applies to coronagraphic \gls{tdi}, which therefore has to be extended to second generation.
    
    \item \textit{Signal superposition:} \gls{lisa} data are expected to be signal dominated.
    As an order of magnitude, we foresee $10^7$ \glspl{gb} to be simultaneously present in the data~\cite{colpi_lisa_2024}.
    As a consequence, it is important to understand how coronagraphic \gls{tdi} behaves when faced with more than a single point-like \gls{gw} source.
    
    \item \textit{Analysis in the presence of instrumental noise:} Our previous work provided a noise model but lacked a systematic study of \gls{gw} sources in the presence of instrumental noise.
    Such a study allows one to determine a \gls{snr} threshold for sky localization with coronagraphic \gls{tdi} which is a fundamental step to assess the applicability of the method.
\end{enumerate}
We address each of these points in this paper, grounding coronagraphic \gls{tdi} in more realistic scenarios and paving the way for concrete applications.

%--------------------------------------
% Second generation TDI
%--------------------------------------
\section{Second-generation TDI} \label{sec:tdi2}

In this section, we recall the definition and principle of coronagraphic \gls{tdi} before extending it.
We then derive an analytic noise model specific to this variable which we validate against instrument simulations.
Finally, we compare the sky-localization capabilities of first- and second-generation combinations.

% Principle
%--------------------------------------
\subsection{Principle} \label{sec:tdi2-principle}

% Definition
In the frequency domain, coronagraphic \gls{tdi} is a linear combination of the first-generation Sagnac variables ${(\tilde{\alpha}, \tilde{\beta}, \tilde{\gamma})}$, where we use the tilde notation to denote the Fourier transform.
The Sagnac combination $\tilde{\alpha}$ is not to be confused with the right ascension $\alpha$.
As a linear combination, coronagraphic \gls{tdi} can be written as 
\begin{equation}
    \tilde{\kappa}(f, \delta, \alpha) = \vec{A}(f, \delta, \alpha) \cdot \vec{D}(f), \label{eq:kappa}
\end{equation}
as in {Eq.\,(12)} in~\cite{costa_barroso_coronagraphic_2025}, with ${\vec{D}(f) = (\tilde{\alpha}(f), \tilde{\beta}(f), \tilde{\gamma}(f))^T}$.
This expression separates $\tilde{\kappa}$ into two conceptually distinct components:
\begin{itemize}
    \item \textit{Geometric response:} The vector $\vec{A}(f, \delta, \alpha)$ which contains the \textit{coronagraphic coefficients}.
    These coefficients take advantage of the geometric response of \gls{lisa} to \glspl{gw} to ensure $\tilde{\kappa}$'s \textit{coronagraphic property}.
    This is achieved by taking both the geometric response of the instrument, i.e., the antenna pattern functions, and the geometric response at the \gls{tdi} level into account.
    Both have to be accounted for because \gls{tdi} relies on the application of delay operators which depend on the position of the spacecraft and, as a consequence, propagates information in a specific way which also depends on the geometric response of \gls{lisa}.
    \item \textit{Data stream:} The vector $\vec{D}(f)$ which carries the data.
    The sky dependence and coronagraphic property of $\tilde{\kappa}$ come from the coefficients in ${\vec{A}(f, \delta, \alpha)}$.
    However, $\tilde{\kappa}$ still depends on the standard \gls{tdi} variables through ${\vec{D}(f)}$.
    As a consequence, $\tilde{\kappa}$ faces the same limitations as ${(\tilde{\alpha},\tilde{\beta},\tilde{\gamma})}$.
\end{itemize}

% Problem and solution
A major limitation of first-generation variables arises when spacecraft motion is introduced~\cite{tinto_time_2004}.
In this scenario, only second-generation variables achieve the sensitivity required by data analysis~\cite{PhysRevD.105.062006}.
Given the clear separation between $\vec{A}(f, \delta, \alpha)$ and $\vec{D}(f)$ in \cref{eq:kappa}, one way to tackle this problem from the coronagraphic perspective is to simply replace the first-generation vector ${\vec{D}(f)}$ by a second-generation vector ${\vec{D}_2(f) = (\tilde{\alpha}_2, \tilde{\beta}_2, \tilde{\gamma}_2)^T}$.
Then, \cref{eq:kappa} becomes
\begin{equation}
    \tilde{\kappa}_2(f, \delta, \alpha) = \vec{A}(f, \delta, \alpha) \cdot \vec{D}_2(f), \label{eq:kappa2}
\end{equation}
which defines the second-generation coronagraphic variable $\tilde{\kappa}_2$.
Because laser frequency noise is reduced to sufficient levels in $\vec{D}_2(f)$, we know this will also hold for $\tilde{\kappa}_2$.
What remains unclear from \cref{eq:kappa2} is whether or not the signal-canceling property of $\tilde{\kappa}_2$ is preserved.

% Argument
The argument given in~\cite{tinto_lisa_2004} is based on the fact that first- and second-generation variables are related by
\begin{equation}
    \alpha_2 (t) \approx \alpha(t) - \alpha(t - L_1 - L_2 - L_3), \label{eq:tdi12}
\end{equation}
where $L_i$, with $i \in \{1, 2, 3\}$, denotes the light travel time along arm $i$ (see~\cite{costa_barroso_coronagraphic_2025,tinto_lisa_2004} for details on labeling conventions).
In the frequency domain, \cref{eq:tdi12} implies that the two variables just differ by a common factor.
As a consequence, the coronagraphic property of $\tilde{\kappa}_2$ should not be affected by this overall factor.
Therefore, in principle, $\tilde{\kappa}_2$ should maintain its ability to localize a point-like \gls{gw} in the sky.

% Time domain
This approach also naturally extends to the time domain.
As shown in~\cite{tinto_lisa_2004}, after deriving coronagraphic \gls{tdi} in the frequency domain one can go back to the time domain by applying an inverse Fourier transform.
This results in $\kappa$ being a sum of delayed and amplified Sagnac variables ${(\alpha, \beta, \gamma)}$ in the time domain.
Its expression is given in Eq.\,(A24) in~\cite{costa_barroso_coronagraphic_2025}
\begin{align}
    \kappa(t, \delta, \alpha) &= \sum_{k=1}^{27} [ A_1^{(k)}\alpha(t+\Delta_1^{(k)}) + A_2^{(k)}\beta(t+\Delta_2^{(k)}) \nonumber \\
    & \quad + A_3^{(k)}\gamma(t+\Delta_3^{(k)})], \label{eq:time}
\end{align}
with exact definitions of the amplitudes $A_i^{(k)}$ and delays $\Delta_i^{(k)}$ for ${i \in \{1,2,3\}}$ and ${k \in \{1, \dots, 27\}}$ provided in~\cite{costa_barroso_coronagraphic_2025,tinto_lisa_2004}.
Just as ${\vec{A}(f, \delta, \alpha)}$ is unchanged in the frequency domain, we apply the same amplitudes and delays to second-generation variables ${(\alpha_2, \beta_2, \gamma_2)}$ in the time domain, leading to:
\begin{align}
    \kappa_2(t, \delta, \alpha) &= \sum_{k=1}^{27} [ A_1^{(k)}\alpha_2(t+\Delta_1^{(k)}) + A_2^{(k)}\beta_2(t+\Delta_2^{(k)}) \nonumber \\
    & \quad + A_3^{(k)}\gamma_2(t+\Delta_3^{(k)})].
\end{align}
Although not explicitly written in this equation, the sky dependence of $\kappa_2$ comes from the dependence of amplitudes and time shifts on ${(\delta, \alpha)}$.
In this paper, we use both the time- and frequency-domain implementations of first- and second-generation coronagraphic \gls{tdi}.

% Noise model
%--------------------------------------
\subsection{Noise model} \label{sec:tdi2-noise}

% Definition
We derive a noise model for $\tilde{\kappa}_2$ in the frequency domain, following the reasoning already outlined in~\cite{costa_barroso_coronagraphic_2025} for $\tilde{\kappa}$.
The key result from~\cite{costa_barroso_coronagraphic_2025} is {Eq.\,(19)}, which can also be written as
\begin{align}
    \begin{split}
        S_{\kappa\kappa}(f, \delta, \alpha) & = \langle \tilde{\kappa}(f, \delta, \alpha) \tilde{\kappa}^*(f, \delta, \alpha) \rangle \\
        & = \vec{A}^{\,T}(f, \delta, \alpha) \cdot S(f) \cdot \vec{A}^\ast(f, \delta, \alpha).
        \label{eq:noise}
    \end{split}
\end{align}
In~\cite{costa_barroso_coronagraphic_2025}, $S$ is the covariance matrix for the \textit{first-generation} Sagnac variables such that
\begin{equation}
    S(f) =
    \begin{pmatrix}
        S_{\alpha\alpha} & S_{\alpha\beta} & S_{\alpha\gamma} \\
        S_{\beta\alpha} & S_{\beta\beta} & S_{\beta\gamma} \\
        S_{\gamma\alpha} & S_{\gamma\beta} & S_{\gamma\gamma}
    \end{pmatrix}. \label{eq:cov}
\end{equation}
For the \textit{second-generation} coronagraphic \gls{tdi} combination $\tilde{\kappa}_2$, $S$ has to be updated.

% Covariance matrix
To obtain the matrix elements in \cref{eq:cov}, we apply specific \gls{tdi} transfer functions to the physically motivated spectral shapes of the two dominant secondary noise sources, namely \gls{tm} acceleration and \gls{oms} noise.
Assuming noise amplitudes to be identical in all spacecraft, these spectral shapes are given by~\cite{lisa2017}:
\begin{align}
    \begin{split}
    S^{\mathrm{TM}}(f) &= A^2 \times \Big[1+\Big(\frac{0.4 \times 10^{-3} \mathrm{Hz}}{f}\Big)^2\Big] \\
    &\quad \times \Big[1+\Big(\frac{f}{8 \times 10^{-3} \mathrm{Hz}}\Big)^4\Big] \\
    &\quad \times \Big(\frac{1}{\omega c}\Big)^2 \times (\mathrm{m}^2/\mathrm{s}^3)
    \end{split} \label{eq:tm_shape} \\[1ex]
    \begin{split}
    S^{\mathrm{OMS}}(f) &= P^2 \times \Big[1+\Big(\frac{2 \times 10^{-3} \mathrm{Hz}}{f}\Big)^2\Big] \\
    &\quad \times \Big(\frac{\omega}{c}\Big)^2 \times (\mathrm{m}^2/\mathrm{Hz}),
    \end{split} \label{eq:spectral_shape}
\end{align}
where ${\omega = 2 \pi f}$.
The amplitudes $A$ and $P$ used in this work are given in \cref{tab:noise-sims} under \gls{tm} acceleration and \gls{oms} noise \gls{asd} respectively.
The suitable \gls{tdi} transfer functions are computed using the methodology introduced in~\cite{PhysRevD.108.082004}.
For \textit{second-generation} Sagnac variables and assuming \textit{equal arm lengths}, we have
\begin{align}
    \begin{split}
    S_{\alpha\alpha} &= S_{\beta\beta} = S_{\gamma\gamma} \\
    &= 24 \times \sin^2(3 \omega L / 2) \times S^{\mathrm{OMS}} \\
    &\quad + 8 \times [3 - 2\cos(\omega L) - \cos(3\omega L)] \\
    &\qquad \times [1 - \cos(3 \omega L)] \times S^{\mathrm{TM}}
    \end{split} \label{eq:psd} \\[1ex]
    \begin{split}
    S_{\alpha\beta} &= S_{\alpha\gamma} = S_{\beta\gamma} \\
    &= 4 \times [2\cos(\omega L) + \cos(2 \omega L)] \\
    &\qquad \times [1 - \cos(3 \omega L)] \times S^{\mathrm{OMS}} \\
    &\quad - 8 \times [\cos(\omega L) - \cos(2 \omega L)]^2 \times S^{\mathrm{TM}}
    \end{split} \label{eq:csd}
\end{align}
for the \glspl{psd} $S_{\alpha\alpha}$, $S_{\beta\beta}$ and $S_{\gamma\gamma}$, and for the \glspl{csd} $S_{\alpha\beta}$, $S_{\alpha\gamma}$ and $S_{\beta\gamma}$.

% Limitations
Paired with the coronagraphic coefficients ${\vec{A}(f,\delta,\alpha)}$, this covariance matrix ${S(f)}$ provides us with a simple analytic noise model for $\tilde{\kappa}_2$ assuming identical noise levels in all spacecraft and equal arm lengths.
This model can be made more complex by relaxing some of these assumptions, e.g., considering unequal arm lengths.
Such modifications would impact just the covariance matrix $S(f)$, not the coefficients ${\vec{A}(f, \delta, \alpha)}$.
These extensions are an interesting direction for future investigations, but this simple analytic model serves the purpose of the analyses presented in the rest of this paper (see Sec.~\ref{sec:tdi2-validation}).

% AET
At this stage, it is worth mentioning that coronagraphic \gls{tdi}, in first or second generation, can be easily extended to the ${(\mathcal{A}, \mathcal{E}, \mathcal{T})}$ basis.
This basis is the Sagnac equivalent of the usual ${(A, E, T)}$ quasi-orthogonal basis constructed from the Michelson variables ${(X, Y, Z)}$.
Such a basis is designed to diagonalize the covariance matrix ${S(f)}$, simplifying data analysis~\cite{PhysRevD.66.122002}.
This is outside of the scope of this paper and is presented in Appendix~\ref{sec:AET}.

% Table
\begin{table*}[tp]
\begin{ruledtabular}
\begin{tabular}{c c c c c}
 & Simulation A & Simulation B & Simulation C & Simulation D\\
\hline
Orbits & Static (equal arms) & Kepler & Kepler & Static (unequal arms) \\
Sampling frequency [Hz] & 4 & 4 & 4 & 4 \\
Duration [days] & 10 & 10 & 5 & 5 \\
Central frequency [Hz] & ${2.816 \times 10^{14}}$ & ${2.816 \times 10^{14}}$ & ${2.816 \times 10^{14}}$ & ${2.816 \times 10^{14}}$ \\
\acrshort{tm} acceleration noise \acrshort{asd} [m$\cdot\mathrm{s}^{-2}/\sqrt{\mathrm{Hz}}$] & ${2.4 \times 10^{-15}}$ & ${2.4 \times 10^{-15}}$ & ${2.4 \times 10^{-15}}$ & ${2.4 \times 10^{-15}}$ \\
\acrshort{oms} noise \acrshort{asd} [m/$\sqrt{\mathrm{Hz}}$] & ${6.35 \times 10^{-12}}$ & ${6.35 \times 10^{-12}}$ & ${6.35 \times 10^{-12}}$ & ${7.9 \times 10^{-12}}$ \\
Laser frequency noise \acrshort{asd} [Hz/$\sqrt{\mathrm{Hz}}$] & 0 & 30 & 30 & 30 \\
\end{tabular}
\end{ruledtabular}
\caption{\label{tab:noise-sims}Simulation parameters used in instrumental noise simulations. Note that we assume the same \gls{asd} across all six \glspl{mosa}. Also note that the \gls{asd} values, laser frequency noise level and central frequency used here are the default values from \textsc{LISA Instrument}. Both values used for \gls{oms} noise are defaults of \textsc{LISA Instrument}, but for different \glspl{mosa}, as we set the same level in all \glspl{mosa}, this is irrelevant. For the unequal-arm static constellation used in simulation D, we also use the default arm lengths from \textsc{LISA Orbits}.}
\end{table*}

% Validation
%--------------------------------------
\subsection{Validation} \label{sec:tdi2-validation}

% Chapeau
Both $\tilde{\kappa}$ and $\tilde{\kappa}_2$ are implemented in a Python package, \textsc{NullStream}, available at \href{https://gitlab.in2p3.fr/costa/null-stream-tdi}{https://gitlab.in2p3.fr/costa/null-stream-tdi}.
This package is integrated with other software tools commonly used to simulate \gls{lisa} data, namely \textsc{LISA Orbits}~\cite{orbits}, \textsc{LISA GW Response}~\cite{gwresponse}, \textsc{LISA Instrument}~\cite{lisainstrument} and \textsc{PyTDI}~\cite{pytdi}.
To validate the second-generation coronagraphic \gls{tdi} variable $\tilde{\kappa}_2$ we proceed in two steps:
\begin{enumerate}
    \item First, we show that $\tilde{\kappa}_2$ reduces laser frequency noise in instrument simulations generated with moving spacecraft.
    \item Second, we show that $\tilde{\kappa}_2$ preserves its sky-localization property despite using the coefficients $\vec{A}(f, \delta, \alpha)$ originally computed for $\tilde{\kappa}$.
\end{enumerate}

% Laser frequency noise reduction
%--------------------------------------
\subsubsection{Laser frequency noise reduction}

% Simulation
We run two simulations with \textsc{LISA Orbits} and \textsc{LISA Instrument}, as all instrument simulations presented in this paper.
Simulation parameters are reported in \cref{tab:noise-sims} under simulation A and B.
The two simulations differ by the orbits used and the presence (or absence) of laser frequency noise, while both simulations have a duration of 10\,days, which is enough to capture the low-frequency part of the noise spectrum (between $10^{-4}$ to $10^{-3}$\,Hz).
The purpose of the first simulation, simulation A, is to validate the analytic noise model \cref{eq:noise}.
This simulation assumes static equal-arm-length orbits and no laser frequency noise, thus matching the assumptions of the analytic noise model.
By contrast, simulation B shifts away from these assumptions by introducing both laser frequency noise and dynamic unequal-arm-length orbits.
In this case, we expect the second-generation variable $\tilde{\kappa}_2$ to be able to reduce laser frequency noise, unlike $\tilde{\kappa}$.
For both simulations, we compute both first- and second-generation coronagraphic \gls{tdi} in the time domain at an arbitrary sky position ${(\delta_0=0.86,\alpha_0=4.38)\,\mathrm{rad}}$.
The same conclusions could be derived from any position outside of the plane of the constellation where coronagraphic \gls{tdi} is exactly zero, as discussed in~\cite{costa_barroso_coronagraphic_2025}.
We then compute the \gls{psd} of both the $\kappa$ and $\kappa_2$ time series using Welch's method~\cite{welch}.
To avoid reaching a numerical noise floor while in band, we apply a low-pass filter to the Sagnac basis before computing $\kappa$ and $\kappa_2$.
The resulting \glspl{psd} are finally compared against the analytic noise models proposed in Sec.~\ref{sec:tdi2-noise}.

% Figure: PSDs
\begin{figure}[tbp]
    \centering
    % TDI 1
    \begin{subfigure}[t]{\columnwidth}
        \centering
        \includegraphics[width=0.96\linewidth]{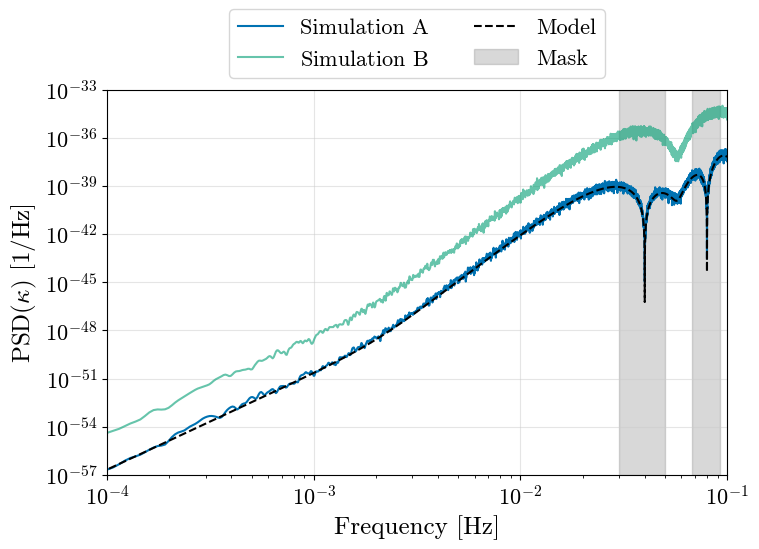}
        \caption{\label{fig:psd-tdi1}Noise spectrum for first-generation coronagraphic \gls{tdi}.}
    \end{subfigure}
    \vspace{0.3cm}
    % TDI 2
    \begin{subfigure}[t]{\columnwidth}
        \centering
        \includegraphics[width=0.96\linewidth]{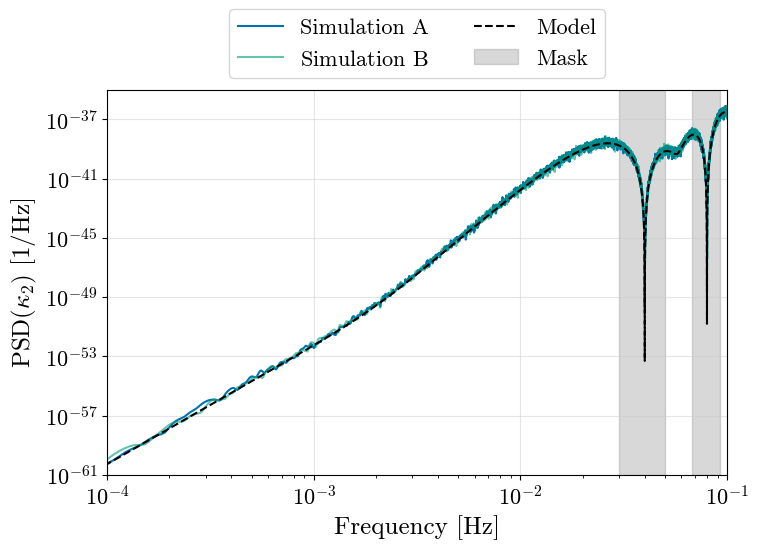}
        \caption{\label{fig:psd-tdi2}Noise spectrum for second-generation coronagraphic \gls{tdi}.}
    \end{subfigure}

    \caption{\label{fig:psds}Noise spectra for first- and second-generation coronagraphic \gls{tdi} obtained from simulation A and B (see \cref{tab:noise-sims}) at ${(\delta_0 = 0.86,\alpha_0 = 4.38)}$\,rad.
    An order-10 zero-phase Butterworth low-pass filter with cutoff frequency $0.3$\,Hz was applied to the Sagnac variables before computing $\kappa$ and $\kappa_2$.
    The \glspl{psd} were computed with Welch's method with a segment length of $2^{19}$ points (approximately $1.5$\,days), a Kaiser window (with shape parameter $40$) and no detrending.
    Analytic models take \gls{tm} acceleration and \gls{oms} noise into account, assuming equal levels across \glspl{mosa} and equal arms.}
\end{figure}

\begin{figure}[tbp]
    \centering
    % Relative error
    \begin{subfigure}[t]{\columnwidth}
        \centering
        \includegraphics[width=0.96\linewidth]{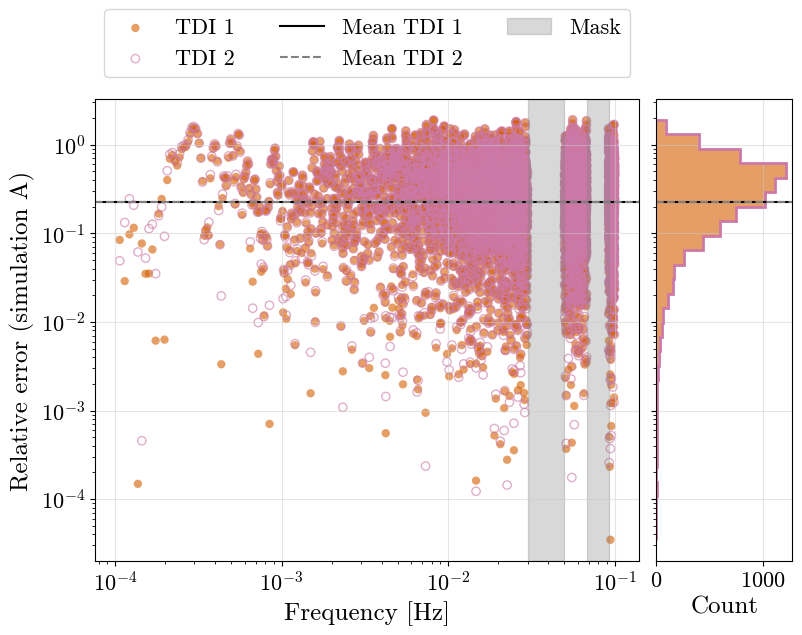}
        \caption{\label{fig:relative-error}Comparison between simulation results and analytic noise model for both first- and second-generation coronagraphic \gls{tdi}.
        Simulation A was generated under the same assumptions as the noise model (see \cref{tab:noise-sims}).
        The \gls{tdi} 1 histogram has mean $2.27 \times 10^{-1}$, while the \gls{tdi} 2 histogram has mean $2.28 \times 10^{-1}$.
        Both histograms have standard deviation $3.02 \times 10^{-1}$.}
    \end{subfigure}
    \vspace{0.3cm}
    % Ratio
    \begin{subfigure}[t]{\columnwidth}
        \centering
        \includegraphics[width=0.96\linewidth]{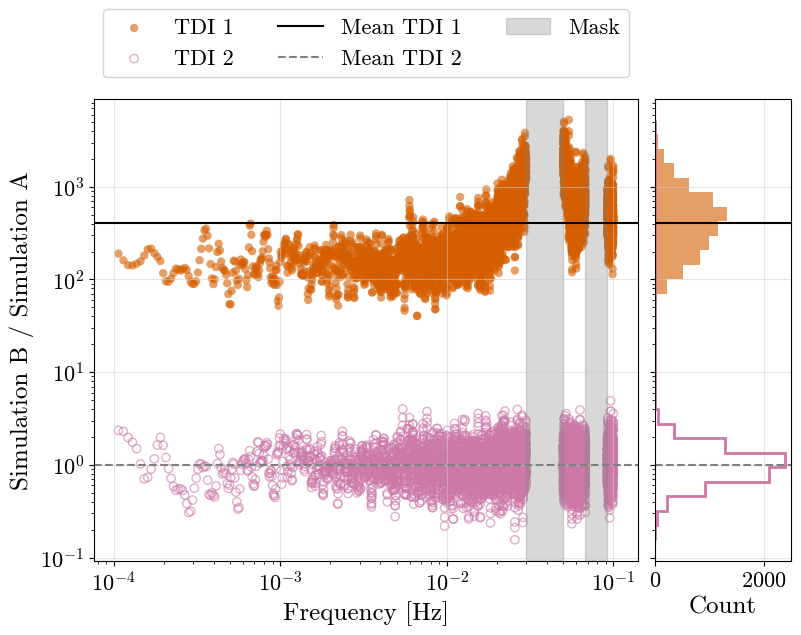}
        \caption{\label{fig:ratio}Comparison between simulation A and simulation B \glspl{psd} obtained for both first- and second-generation coronagraphic \gls{tdi}.
        The \gls{tdi} 1 histogram has mean 406 and standard deviation 492, while the \gls{tdi} 2 histogram has mean $9.93 \times 10^{-1}$ and standard deviation $4.75 \times 10^{-1}$.}
    \end{subfigure}

    \caption{\label{fig:comparisons}Comparisons between \glspl{psd} and noise models presented in \cref{fig:psds}.
    The masked regions correspond to the zeroes in \gls{tdi} transfer functions visible in \cref{fig:psds}.
    These ill-defined regions were not taken into account in the comparisons presented here.
    The means displayed correspond to geometric means.}
\end{figure}

% Results TDI 1
In \cref{fig:psd-tdi1}, we plot the \glspl{psd} of first-generation coronagraphic \gls{tdi} computed from both simulation A and B, together with the associated analytic noise model.
For simulation A, we find a good agreement between the \gls{psd} and the model, with a relative error between the two of the order of $10^{-1}$ (see \cref{fig:relative-error}).
This is consistent with results already presented in~\cite{costa_barroso_coronagraphic_2025}.
By contrast, the \gls{psd} obtained from simulation B, lies well above the analytic noise model.
Because the analytic noise model only describes \gls{oms} and \gls{tm} acceleration noise, this difference indicates the presence of laser frequency noise in $\tilde{\kappa}$.
Indeed, as shown in \cref{fig:ratio}, the \gls{psd} obtained from simulation B is nearly three orders of magnitude larger than the one computed from simulation A.

% Figure: comparisons

% Results TDI 2
In \cref{fig:psd-tdi2}, we plot the results for second-generation coronagraphic \gls{tdi}.
Similar to the first-generation case, we find a good agreement between the \gls{psd} computed from simulation A and the analytic noise model with a relative error between the two of the order of $10^{-1}$ (see \cref{fig:relative-error}).
This result validates the noise model derived in Sec.~\ref{sec:tdi2-noise}.
Importantly, in \cref{fig:psd-tdi2} the \gls{psd} obtained from simulation B also agrees with the noise model.
Indeed, the ratio between simulation A and B is of the order of 1 for second-generation \gls{tdi}, instead of $10^3$ in the first-generation case (see \cref{fig:ratio}).
This means that the noise present in $\kappa_2$ has been reduced to \gls{tm} acceleration and \gls{oms} noise, demonstrating the effectiveness of $\kappa_2$.
This result also justifies the usage of the simple analytic noise model derived in Sec.~\ref{sec:tdi2-noise} for simulations as complex as simulation B.

% Coronagraphic property
%--------------------------------------
\subsubsection{Coronagraphic property}

% Simulation
We run a simulation of a \gls{gb} generated with \textsc{LISA Orbits}, \textsc{LISA Instrument} and \textsc{LISA GW Response}, as all \gls{gb} simulations reported in this paper.
We choose this type of source for convenience: because \glspl{gb} are quasi-monochromatic sources in the \gls{lisa} band, they are well-localized in the frequency domain near their characteristic frequency $f_0$.
The parameters of the simulated source are reported in \cref{tab:sim-params-gb}.
These parameters are arbitrary, except for the amplitude which is chosen to be unrealistically high for clarity (detectability is discussed later in {Sec.~\ref{sec:applicability}}).
On the instrument simulation side, the parameters used are reported in \cref{tab:noise-sims} under simulation C.
We use similar settings to the more complex instrumental scenario, namely simulation B.
In particular, we introduce the motion of the spacecraft by simulating Keplerian orbits and we include laser frequency noise in addition to \gls{tm} acceleration and \gls{oms} noise.

% Table: GB params
\begin{table}[tbp]
\begin{ruledtabular}
\begin{tabular}{c c c}
Parameter & \gls{gb} & Units \\
\colrule
$\mathcal{A}$ & $5.0 \times 10^{-20}$ & - \\
$f_0$ & 7.97 & mHz \\
$\dot{f}$ & $1.0 \times 10^{-18}$ & Hz/s \\
$\iota$ & 5.39 & rad \\
$\phi$ & 2.76 & rad \\
$\Psi$ & 4.38 & rad \\
$\delta_\star$ & $-1.27$ & rad \\
$\alpha_\star$ & $6.13$ & rad \\
\end{tabular}
\end{ruledtabular}
\caption{\label{tab:sim-params-gb}Parameters of arbitrary \gls{gb} injection used to generate \cref{fig:sky}.
The parameters $\{\Psi, \delta_\star, \alpha_\star\}$ are given in the \gls{icrs} frame, following \textsc{LISA GW Response}'s convention.
The data are simulated at a sampling rate of 4\,Hz for 5\,days ($1.728 \times 10^{6}$ data points).}
\end{table}

\begin{figure}[tbp]
    \centering
    % 1st generation
    \begin{subfigure}[t]{\columnwidth}
        \centering
        \includegraphics[width=0.96\linewidth]{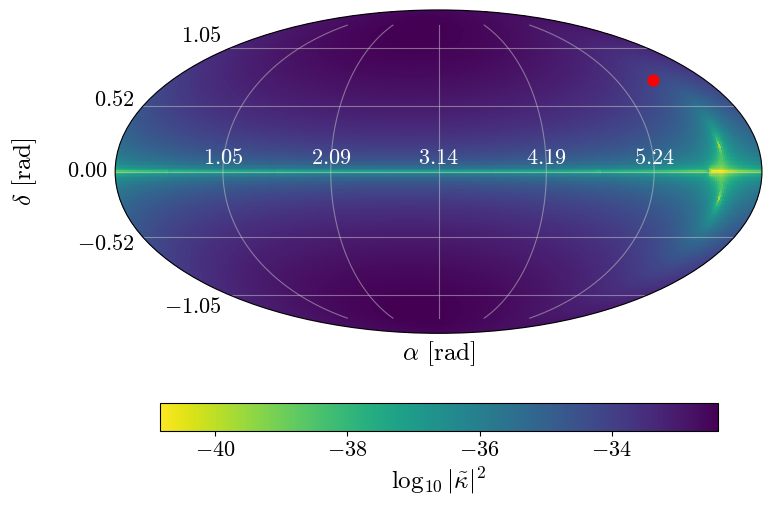}
        \caption{\label{fig:sky-kappa-1}First-generation coronagraphic \gls{tdi} sky map.}
    \end{subfigure}
    \vspace{0.3cm}
    % 2nd generation
    \begin{subfigure}[t]{\columnwidth}
        \centering
        \includegraphics[width=0.96\linewidth]{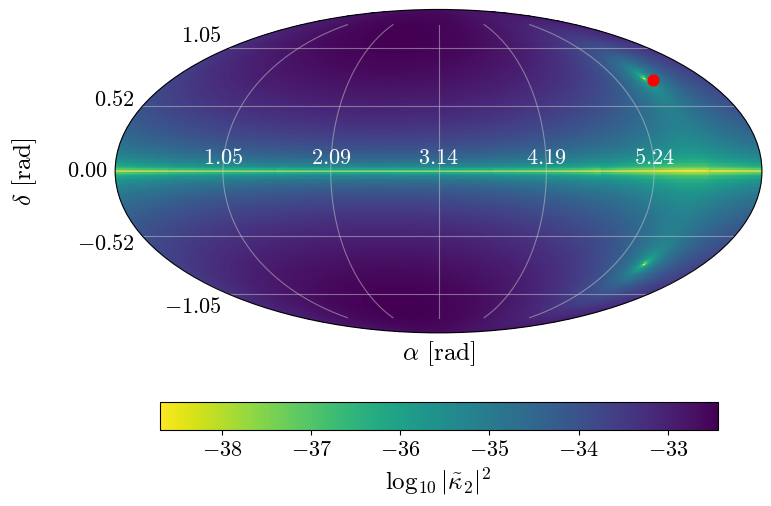}
        \caption{\label{fig:sky-kappa-2}Second-generation coronagraphic \gls{tdi} sky map.}
    \end{subfigure}

    \caption{\label{fig:sky}Coronagraphic \gls{tdi} sky maps in the \gls{lisa} frame. The maps were computed at ${f_0 = 7.97}$\,mHz on a rectangular ${300 \times 300}$ grid, corresponding to a pixel size of $0.7\,\mathrm{deg}^2$ at the equator. The red dot represents the position of the source in the \gls{lisa} frame.}
\end{figure}

% Results
To visualize the sky-localization performance of both first- and second-generation coronagraphic variables, we display the sky map of both $|\tilde{\kappa}|^2$ and $|\tilde{\kappa}_2|^2$ at $f_0$ in the \gls{lisa} frame in \cref{fig:sky}.
In both first-generation (\cref{fig:sky-kappa-1}) and second-generation (\cref{fig:sky-kappa-2}) maps, we see two minima together with the known zero structure at $\delta=0$.
More precisely, the minimum in the northern hemisphere in \cref{fig:sky-kappa-1} has an angular separation of {44\,deg} to the position of the source ${(\delta_\star, \alpha_\star)}$ in the \gls{lisa} frame.
The same quantity takes value $3$\,deg for the second-generation combination shown in \cref{fig:sky-kappa-2}, which confirms that $\tilde{\kappa}_2$ preserves the coronagraphic property.
This difference in sky-localization performance highlights the benefit of using $\tilde{\kappa}_2$ over $\tilde{\kappa}$.
Regarding the $3$\,deg angular separation obtained in the second-generation case, we do expect noise to introduce some bias and discuss this effect in more detail in {Sec.~\ref{sec:applicability}}.

% Figure: Coronagraphic property

% Conclusion
%--------------------------------------
\subsection{Conclusion} \label{sec:tdi2-conclusion}

The results presented in this section show that $\tilde{\kappa}_2$ is the suitable tool to analyze data in more complex scenarios including motion of the spacecraft and instrumental noise.
For this reason, we use it in the rest of this paper whenever instrumental noise is considered (see Sec.~\ref{sec:analysis}).
Although $\tilde{\kappa}_2$ represents a significant improvement compared to $\tilde{\kappa}$, the limitations related to the geometric response term ${\vec{A}(f, \delta, \alpha)}$ remain.
Specifically, the coefficients ${\vec{A}(f, \delta, \alpha)}$ are computed for a frozen \gls{lisa} configuration and do not follow the motion of the constellation.
This was already mentioned in~\cite{costa_barroso_coronagraphic_2025} and restricts coronagraphic \gls{tdi} to the analysis of short time segments of the order of weeks.
The simulations analyzed in this paper are even shorter, of just a few days, and therefore should not be significantly impacted by by the mismatch between the frozen coefficients ${\vec{A}(f, \delta, \alpha)}$ and the time-evolving position of the spacecraft in the data.
Indeed, from the case studied in Sec.~\ref{sec:tdi2-validation}, this only introduced a bias of $3$\,deg which can also be coupled to instrumental noise.
The most limiting implication of using a short time window is the small amount of \gls{snr} accumulated (see Sec.~\ref{sec:analysis}).
Therefore, deriving time-dependent coronagraphic coefficients is an interesting extension which should be considered in the future.

%--------------------------------------
% Signal superposition
%--------------------------------------
\section{Signal superposition} \label{sec:superposition}

In this section, we focus on the problem of signal superposition without instrumental noise.
We first analyze a set of simulations of pairs of \glspl{gb} which serve as toy models to examine $\tilde{\kappa}$'s response to more than a single source.
We then consider the more complex case of multiple \glspl{gb} and a single \gls{mbhb}.

% Toy models
%--------------------------------------
\subsection{Toy models} \label{sec:superposition-toy}

% Intuition
From the definition of coronagraphic \gls{tdi}, we know that for a single point-like \gls{gw} source $\tilde{\kappa}(f, \delta, \alpha)$ tends to zero when $(f, \delta, \alpha) \rightarrow (f_\star, \delta_\star, \alpha_\star)$.
Intuitively, if we consider two quasi-monochromatic sources such as \glspl{gb} with characteristic frequencies $f_0^0$ and $f_0^1$, we expect the coronagraphic property to hold for ${f_0^0 \neq f_0^1}$.
This is because if the two sources do not overlap in the frequency domain, one should be able to treat them independently. 
However, a few scenarios may challenge this reasoning:
\begin{itemize}
    \item \textit{Spatial resolution:} The two \glspl{gb} have different characteristic frequencies (i.e., ${f_0^0 \neq f_0^1}$), but nearby sky positions ${(\delta_\star^0, \alpha_\star^0)}$ and ${(\delta_\star^1, \alpha_\star^1)}$.
    In this case, we want to understand what happens when two sources overlap in space, but not in frequency.
    \item \textit{Frequency resolution:} The two \glspl{gb} have close characteristic frequencies (i.e., ${f_0^0 \approx f_0^1}$), but  different sky positions.
    In this case, we want to see what happens when two sources overlap in frequency, but not in space.
    \item \textit{Relative amplitude:} The two \glspl{gb} have identical characteristic frequencies (i.e., ${f_0^0=f_0^1}$), different sky positions and significantly different amplitudes.
    In this case, we want to check whether the louder source can still be localized despite both \glspl{gb} having the same characteristic frequency.
\end{itemize}
In this section, we test each of these scenarios using three sets of noiseless simulations.

% Method
All three datasets are analyzed with the same method:
\begin{enumerate}
    \item Compute two sky maps $\tilde{\kappa}(f_{0}^0, \delta, \alpha)$ and $\tilde{\kappa}(f_{0}^1, \delta, \alpha)$, similar to \cref{fig:sky-kappa-1}.
    We use first-generation coronagraphic \gls{tdi} for this analysis because no instrumental noise is included in the simulations considered in this section.
    \item Identify the position ${(\delta_{min}^i, \alpha_{min}^i)}$ of the minimum of sky map $|\tilde{\kappa}(f_{0}^i, \delta, \alpha)|^2$, for $i \in \{0,1\}$.
    \item Compare the position of the minimum to the known position of the source, through angular separation
    \begin{equation}
    \begin{split}
    \Delta \sigma^i &= \arccos[\sin(\delta_{min}^i) \sin(\delta_\star^i) \\
    &\quad + \cos(\delta_{min}^i) \cos(\delta_\star^i) \cos(\alpha_{min}^i - \alpha_\star^i)].
    \end{split}
    \label{eq:angular-separation}
    \end{equation}
\end{enumerate}
The goal is to quantify with $\Delta \sigma^i$ how much the presence of a second source biases the estimated position of each \gls{gw} source in the three scenarios.

% Spatial resolution
%--------------------------------------
\subsubsection{Spatial resolution}

% Table: spatial resolution
\begin{table}[tbp]
\begin{ruledtabular}
\begin{tabular}{c c c c c}
Parameter & \gls{gb} 0 & \gls{gb} 1 & Units \\
\colrule
$\mathcal{A}$ & $1.0 \times 10^{-21}$ & $1.0 \times 10^{-21}$ & - \\
$f_0$ & 5.0 & 5.1 & mHz \\
$\dot{f}$ & $1.0 \times 10^{-18}$ & $1.0 \times 10^{-18}$ & Hz/s \\
$\iota$ & 0 & 0 & rad \\
$\phi$ & 0 & 0 & rad \\
$\Psi$ & 0 & 0 & rad \\
$\delta_\star$ & $0.8$ & $[0.640, 1.336]$ & rad \\
$\alpha_\star$ & $2.0$ & $[1.840, 2.336]$ & rad \\
\end{tabular}
\caption{\label{tab:sim-params-two-gb-pos}Simulation parameters used to estimate coronagraphic \gls{tdi}'s ability to localize two sources with similar sky locations but different characteristic frequencies.
The parameters $\{\Psi, \delta_\star, \alpha_\star\}$ are given in the \gls{icrs} frame, following \textsc{LISA GW Response}'s convention.
The data are simulated at a sampling rate of ${0.2}$\,Hz for ${5.8}$\,days ($10^{5}$ data points).
We choose a lower sampling rate than the realistic 4\,Hz to reduce computational cost.}
\end{ruledtabular}
\end{table}

\begin{figure}[tbp]
    \centering
    \includegraphics[width=0.7\linewidth]{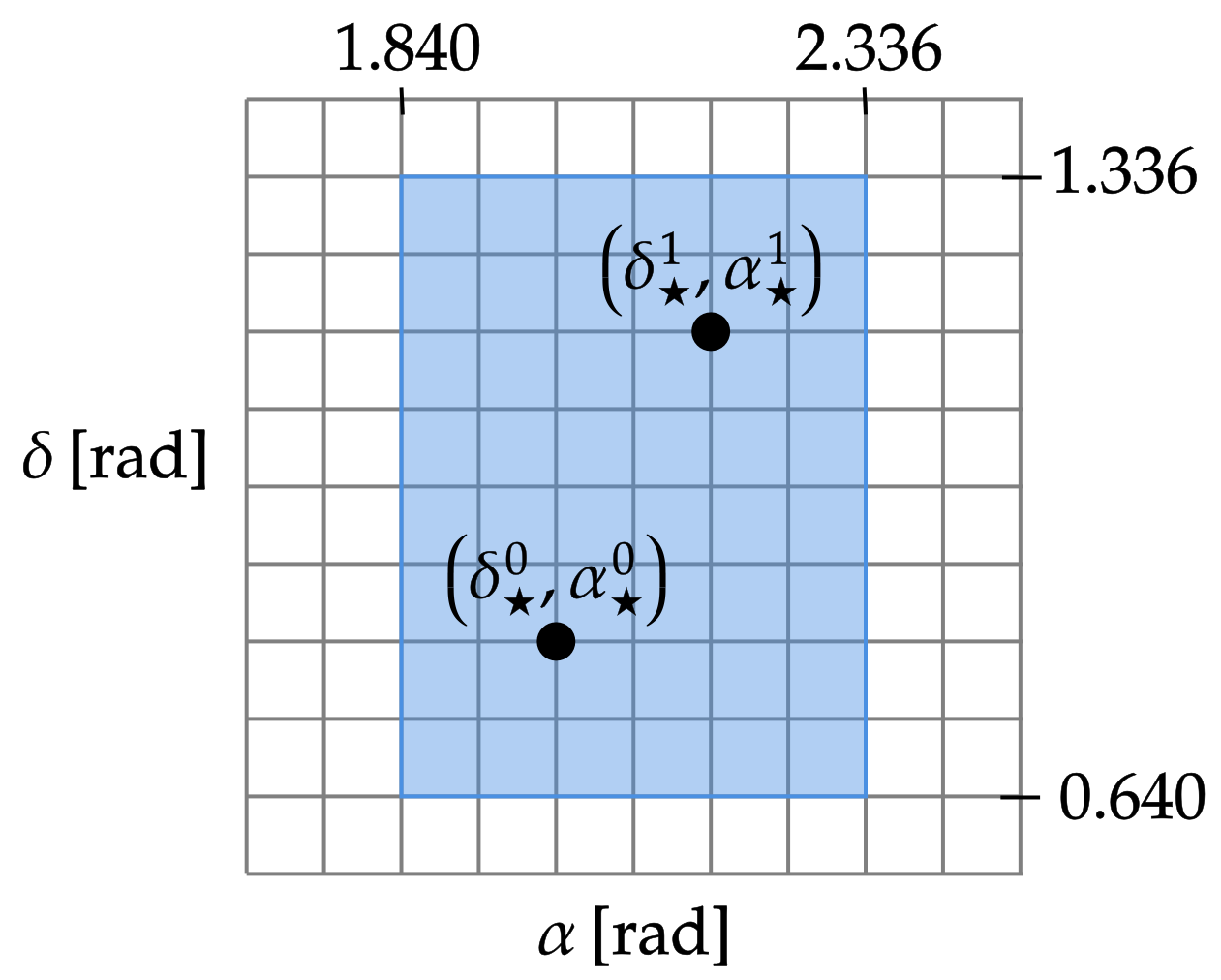}
    \caption{\label{fig:sim-params-two-gb-pos}Schematic representation of the parameter space covered by the \textit{spatial resolution} dataset. All simulations contain two \glspl{gb} at fixed characteristic frequencies $f_0^0$ (\gls{gb} 0) and $f_0^1$ (\gls{gb} 1). The only parameter which varies across simulations is the sky position $(\delta_\star^1, \alpha_\star^1)$ of source 1. The shaded region represents the range of positions covered by the entire dataset.}
\end{figure}

% Simulation
To investigate the first scenario, we generate a set of 1\,024 simulations whose parameters are given in \cref{tab:sim-params-two-gb-pos}.
In this dataset, the frequency difference ${\Delta f = |f_0^0 - f_0^1| = 0.1}$\,mHz remains constant.
Similarly, \gls{gb} 0 is always simulated at the same position ${(\delta_\star^0, \alpha_\star^0)}$.
Simulations differ by the position ${(\delta_\star^1, \alpha_\star^1)}$ of \gls{gb} 1.
As illustrated in \cref{fig:sim-params-two-gb-pos}, source 1 is simulated in a range of positions around ${(\delta_\star^0, \alpha_\star^0)}$.
Except for characteristic frequency and sky position, both sources share exactly the same parameters.

% Figure: spatial resolution diagram

% Results
We analyze this dataset following the method described above with sky maps focused on the region of the two sources (see \cref{tab:sim-params-two-gb-pos}).
Such sky maps are defined by the intervals ${\delta \in [0.4, 1.5]}$\,rad and ${\alpha \in [0.0, 2.5]}$\,rad, corresponding to a pixel size of ${\theta_{pix} = 1.6 \times 10^{-2}}$\,rad.
Across \textit{all} simulations, we obtain ${\Delta \sigma^i < \theta_{pix}}$.
This means that the sky locations estimated through the sky maps $|\tilde{\kappa}(f_0^i, \delta, \alpha)|^2$ were not affected by the presence of another source with different characteristic frequency, within the pixel resolution.
Therefore, sources sharing the same sky position can still be localized independently with coronagraphic \gls{tdi}, provided they are separated in frequency.

% Frequency resolution
%--------------------------------------
\subsubsection{Frequency resolution}

% Table: frequency resolution
\begin{table}[tbp]
\begin{ruledtabular}
\begin{tabular}{c c c c c}
Parameter & \gls{gb} 0 & \gls{gb} 1 & Units \\
\colrule
$\mathcal{A}$ & $1.0 \times 10^{-21}$ & $1.0 \times 10^{-21}$ & - \\
$f_0$ & 5.0 & $[5.0, 5.1]$ & mHz \\
$\dot{f}$ & $1.0 \times 10^{-18}$ & $1.0 \times 10^{-18}$ & Hz/s \\
$\iota$ & 0 & 0 & rad \\
$\phi$ & 0 & 0 & rad \\
$\Psi$ & 0 & 0 & rad \\
$\delta_\star$ & $0.8$ & $1.3$ & rad \\
$\alpha_\star$ & $2.0$ & $0.5$ & rad \\
\end{tabular}
\end{ruledtabular}
\caption{\label{tab:sim-params-two-gb-freq}Simulation parameters used to estimate coronagraphic \gls{tdi}'s ability to localize two sources with different sky locations but similar characteristic frequencies.
The characteristic frequency of the second \gls{gb} ranges from 5 to 5.1\,mHz, with a step of 2\,$\mu$Hz.
The parameters $\{\Psi, \delta_\star, \alpha_\star\}$ are given in the \gls{icrs} frame, following \textsc{LISA GW Response}'s convention.
The data are simulated at a sampling rate of ${0.2}$\,Hz for ${5.8}$\,days ($10^{5}$ data points).
We choose a lower sampling rate than the realistic 4\,Hz to reduce computational cost.}
\end{table}

\begin{figure}[tbp]
    \centering
    \includegraphics[width=0.7\linewidth]{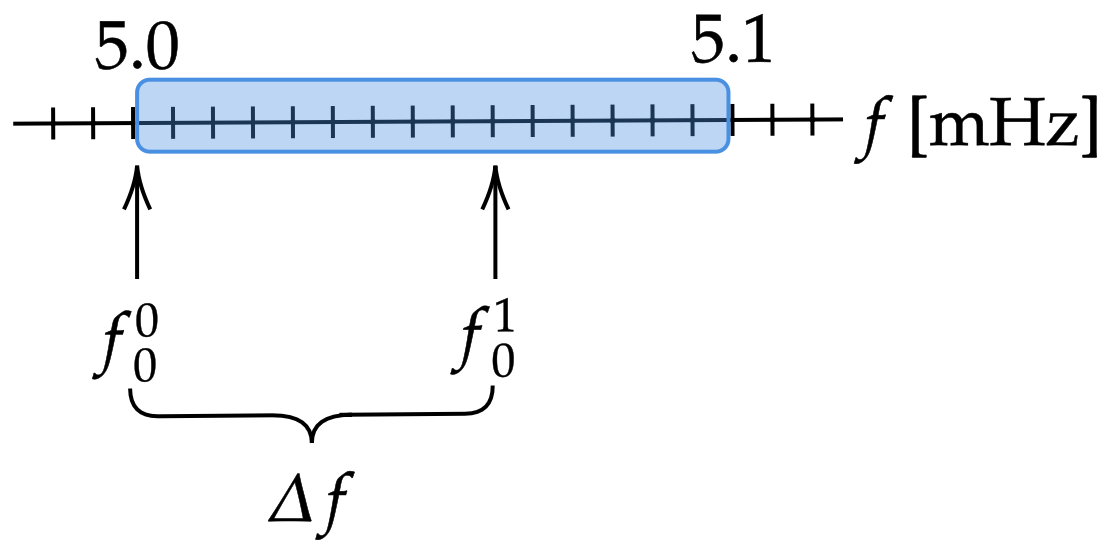}
    \caption{\label{fig:sim-params-two-gb-freq}Schematic representation of the parameter space covered by the \textit{frequency resolution} dataset. All simulations contain two \glspl{gb} at fixed positions $(\delta_\star^0, \alpha_\star^0)$ (\gls{gb} 0) and $(\delta_\star^1, \alpha_\star^1)$ (\gls{gb} 1). The only parameters which vary from one simulation to another are characteristic frequencies $f_0^0$ (\gls{gb} 0) and $f_0^1$ (\gls{gb} 1). The shaded region represents the frequency range covered by the entire dataset.}
\end{figure}

% Simulation
To investigate the second scenario, we simulate a set of 50 simulations whose parameters are given in \cref{tab:sim-params-two-gb-freq}.
Unlike in the previous dataset, the two sources have fixed positions ${(0.8, 2.0)}$\,rad for \gls{gb} 0 and ${(1.3, 0.5)}$\,rad for \gls{gb} 1.
The difference between characteristic frequencies $\Delta f$, however, varies across the dataset.
The first source is fixed at ${f_0^0 = 5.0}$\,mHz, while the characteristic frequency of the second source ranges from {5.0} to {5.1}\,mHz, as illustrated in \cref{fig:sim-params-two-gb-freq}.

% Figure: frequency resolution diagram

% Results
We analyze this dataset using the same sky grid as the one used in the previous dataset.
After analyzing all simulations, we found that down to ${\Delta f = 4\,\mu\mathrm{Hz}}$ we could successfully identify both sources individually.
For \textit{all} simulations satisfying ${\Delta f \geq 4\,\mu\mathrm{Hz}}$, we obtained
\begin{equation}
\begin{cases}
\Delta \sigma^0 &= 2.63 \times 10^{-3}\,\mathrm{rad} \\
\Delta \sigma^1 &= 8.87 \times 10^{-4}\,\mathrm{rad}
\end{cases}
\end{equation}
both of which were well below pixel size ${\theta_{pix} = 1.6 \times 10^{-2}}$\,rad.
Below ${\Delta f = 4\,\mu\mathrm{Hz}}$, given the frequency resolution of the simulation of ${2\,\mu\mathrm{Hz}}$, the two sources effectively overlap in frequency.
Therefore, up to the frequency resolution of the simulation, two sources at distinct frequencies do not disturb one another.

% Relative amplitude
%--------------------------------------
\subsubsection{Relative amplitude}

% Simulation
The results from the analysis of the previous dataset show that overlapping characteristic frequencies are an interesting edge case where coronagraphic \gls{tdi} fails at localizing both \glspl{gb}.
To further investigate this behavior, we generate a set of 80 simulations detailed in \cref{tab:sim-params-two-gb-amplitude}.
In this dataset, both sources share the same characteristic frequency and have fixed positions ${(0.8, 2.0)}$\,rad for \gls{gb} 0 and ${(1.3, 0.5)}$\,rad for \gls{gb} 1.
The parameter which varies across the dataset is the amplitude $\mathcal{A}_1$ of \gls{gb} 1.
The amplitudes $\mathcal{A}_1$ are chosen such that a range of ratios ${\mathcal{A}_1/\mathcal{A}_0}$ going from $10^{-4}$ to 1 (i.e., ${\mathcal{A}_1 \leq \mathcal{A}_0}$) is spanned.
This way, we probe a range of pairs of overlapping sources with different relative loudness.

% Table: relative amplitude
\begin{table}[tbp]
\begin{ruledtabular}
\begin{tabular}{c c c c c}
Parameter & \gls{gb} 0 & \gls{gb} 1 & Units \\
\colrule
$\mathcal{A}$ & $1.0 \times 10^{-21}$ & $[1.0 \times 10^{-25}, 1.0 \times 10^{-21}]$ & - \\
$f_0$ & 5.0 & 5.0 & mHz \\
$\dot{f}$ & $1.0 \times 10^{-18}$ & $1.0 \times 10^{-18}$ & Hz/s \\
$\iota$ & 0 & 0 & rad \\
$\phi$ & 0 & 0 & rad \\
$\Psi$ & 0 & 0 & rad \\
$\delta_\star$ & $0.8$ & $1.3$ & rad \\
$\alpha_\star$ & $2.0$ & $0.5$ & rad \\
\end{tabular}
\caption{\label{tab:sim-params-two-gb-amplitude}Simulation parameters used to estimate coronagraphic \gls{tdi}'s ability to localize two sources with different sky locations, identical characteristic frequency but different relative amplitudes.
The parameters $\{\Psi, \delta_\star, \alpha_\star\}$ are given in the \gls{icrs} frame, following \textsc{LISA GW Response}'s convention.
The data are simulated at a sampling rate of ${0.2}$\,Hz for ${5.8}$\,days ($10^{5}$ data points).
We choose a lower sampling rate than the realistic 4\,Hz to reduce computational cost.}
\end{ruledtabular}
\end{table}

\begin{figure}[tbp]
    \centering
    \includegraphics[width=0.96\linewidth]{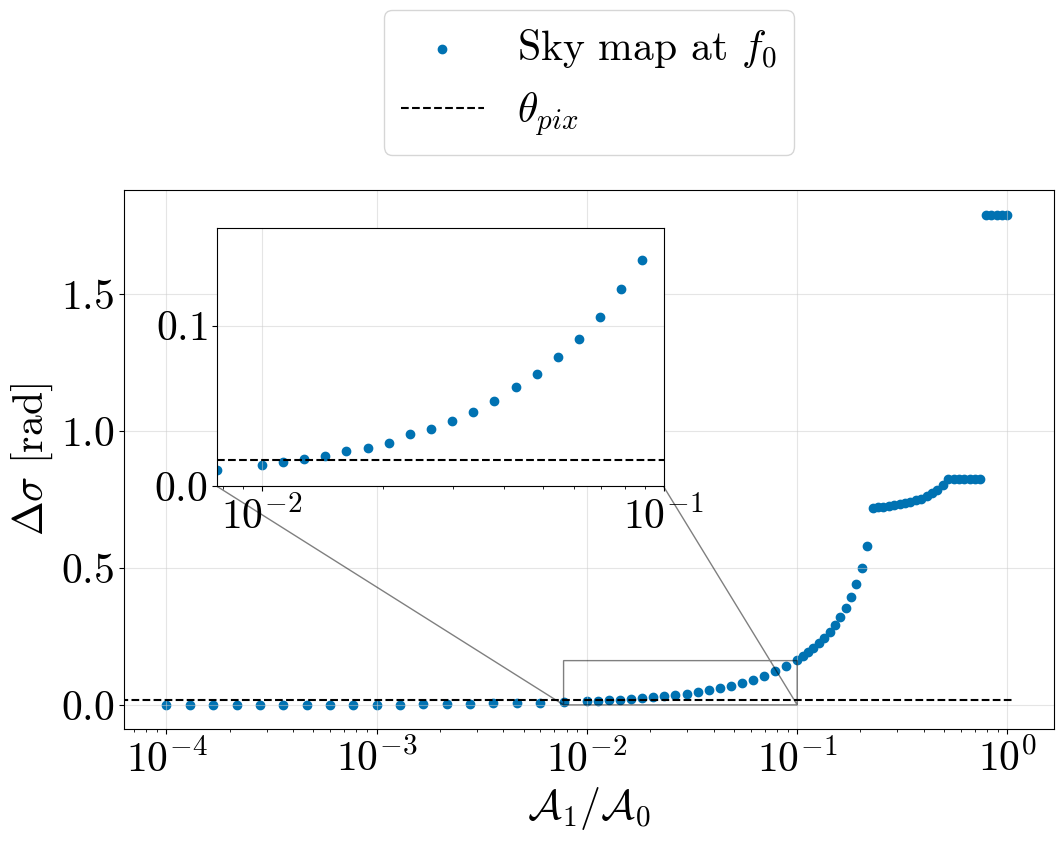}
    \caption{\label{fig:amplitude}Angular separation between estimated and true position of the louder \gls{gb} of amplitude $\mathcal{A}_0$ versus the ratio of the amplitudes of both \glspl{gb}. The inset shows the region where the sky localization of the louder \gls{gb} starts to be significantly affected by the presence of the second \gls{gb}. The dashed line represents the pixel size $\theta_{pix}$ of the sky maps used in this analysis (${\theta_{pix} = 1.6 \times 10^{-2}}$\,rad).}
\end{figure}

% Results
In the analysis of this dataset, we focus on the characteristic frequency $f_0^0$ of the dominant source (recall that ${\mathcal{A}_1 \leq \mathcal{A}_0}$).
For this reason, we simply write ${\Delta \sigma}$ for ${\Delta \sigma^0}$.
Also unlike in previous analyses, we choose a larger sky region defined by the intervals ${\delta \in [0.1, 1.5]}$\,rad and ${\alpha \in [0.0, 6.0]}$\,rad with the usual pixel size ${\theta_{pix} = 1.6 \times 10^{-2}}$\,rad.
This grid covers most of the parameter space so we can capture any important shifts in ${(\delta_{min}^0, \alpha_{min}^0)}$.
The results are plotted in \cref{fig:amplitude} and show that the relative amplitude between two sources affects the sky localization of the louder source.
What we observe is that, as long as \gls{gb} 0 is at least 100 times louder than \gls{gb} 1, the position of \gls{gb} 0 correctly appears on the ${|\tilde{\kappa}(f_0^0, \delta, \alpha)|^2}$ sky map.
As the amplitude $\mathcal{A}_1$ of the quieter source approaches the amplitude $\mathcal{A}_0$ of the louder source, the estimated sky location gets increasingly biased.
When the two sources have comparable amplitudes, no minimum appears in the sky map and the estimated ${(\delta_{min}^0, \alpha_{min}^0)}$ falls near the blind region of $\tilde{\kappa}$, which explains the plateaus of $\Delta \sigma$ for ${\mathcal{A}_1 / \mathcal{A}_0}$ above $10^{-1}$.

% Conclusion
%--------------------------------------
\subsubsection{Conclusion}

The analysis of the datasets presented in this section clarifies how coronagraphic \gls{tdi} responds to data containing more than one source.
After considering three different scenarios, we conclude that:
\begin{itemize}
    \item Provided two \gls{gw} sources do not overlap in frequency, we can retrieve their sky locations by analyzing $|\tilde{\kappa}(f_0^i, \delta, \alpha)|^2$ independently at each of their characteristic frequencies $f_0^i$, even when they share the same sky position.
    \item If two sources overlap in frequency, only the dominant source can be localized, provided it is at least $\sim 100$ times louder than the other.
\end{itemize}
These results establish the limitations of coronagraphic \gls{tdi} when two quasi-monochromatic \gls{gw} sources are present in the data.
Although obtained with \glspl{gb}, these conclusions apply to any pair of point-like sources overlapping in frequency.
In particular, for an \gls{mbhb}, which spans a wide frequency band, we expect sky localization to be affected only at frequencies where another source of comparable amplitude is present.
We test this expectation in the next section.

% Multiple sources
%--------------------------------------
\subsection{Multiple sources} \label{sec:superposition-multiple}

% MBHB + VGB
As a proof of principle, we simulate one \gls{mbhb} from the \gls{ldc} Sangria dataset~\cite{sangria} together with the 49 \glspl{vgb} from~\cite{kupfer2023lisa}.
This is simpler than a realistic \gls{lisa} simulation, but already introduces considerable complexity.
Instead of just two quasi-monochromatic sources, we now have 49 \glspl{vgb} and two types of sources: \glspl{gb} and \glspl{mbhb}.
Specifically, we choose \gls{mbhb}~10, as it was shown to be the source with best resolution in our previous analysis~\cite{costa_barroso_coronagraphic_2025}.
We simulate only the dominant $(2,2)$ mode using the \gls{ldc} software~\cite{ldc}, which implements an IMRPhenomD waveform.
For conciseness, selected simulation parameters are left to Appendix~\ref{sec:params}.

% Method
We analyze this more complex simulation with the same method described in Sec.~\ref{sec:superposition-toy}, but with a few adjustments:
\begin{itemize}
    \item Instead of computing $\tilde{\kappa}(f_0^i, \delta, \alpha)$ on just two frequencies $f_0^i$, we look at a wide range of $1\,000$ frequencies ${f \in [4 \times 10^{-5}, 2 \times 10^{-2}]}$\,Hz.
    This is motivated by the polychromatic nature of the \gls{mbhb} signal which spans multiple frequencies.
    We chose to perform this analysis on $1\,000$ frequencies to reduce computational cost, but this study can easily be extended to more frequencies.
    In addition, we specifically compute $\tilde{\kappa}$ on the characteristic frequencies of the 49 \glspl{vgb} present in the simulation.
    \item We use the wider sky grid used in the \textit{relative amplitude} analysis.
    \item Again, similar to the \textit{relative amplitude} analysis, we focus on the angular separation $\Delta \sigma$ between the position of minimum $|\tilde{\kappa}|^2$ and the position of the source of reference, here \gls{mbhb}~10.
\end{itemize}

% Figure: bias in MBHB sky location
\begin{figure}[tbp]
    \centering
    \includegraphics[width=0.96\linewidth]{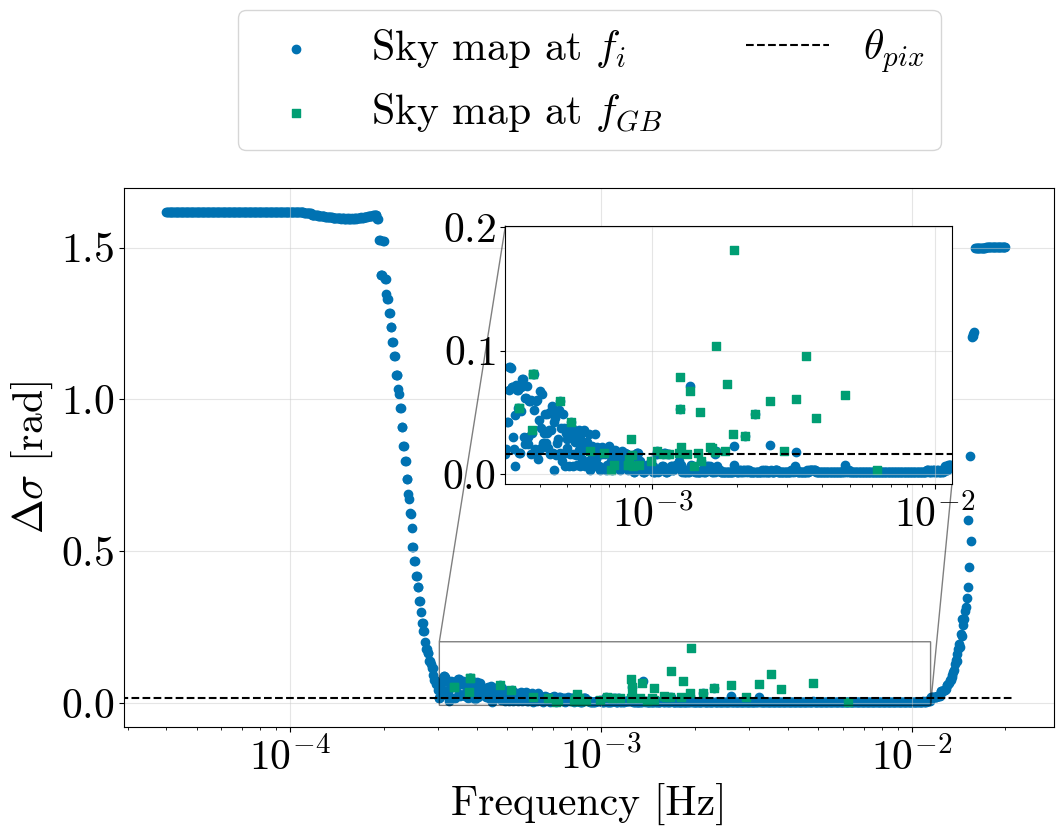}
    \caption{\label{fig:mbhb-vgbs-delta} Angular separation between estimated and true position of \gls{mbhb}~10 in the presence of 49 \glspl{gb}. The inset shows the region corresponding to frequencies spanned by the \gls{mbhb} signal. The dashed line represents the pixel size $\theta_{pix}$ of the sky maps used in this analysis (${\theta_{pix} = 1.6 \times 10^{-2}}$\,rad).}
\end{figure}

% Results - it works overall
The results are shown in \cref{fig:mbhb-vgbs-delta}.
We can distinguish two regimes in \cref{fig:mbhb-vgbs-delta}.
There is a region below about ${3 \times 10^{-4}}$ and above about $10^{-2}$\,Hz where ${\Delta \sigma \geq 1.5}$\,rad.
This corresponds to the region outside of the frequency range of \gls{mbhb}~10.
This is a region without any \gls{gw} signal where the minimum of $\tilde{\kappa}$ falls into the blind region around the equator (see \cref{fig:sky-kappa-1}).
On the other hand, the angular separation $\Delta \sigma$ is much smaller in the complementary region, which corresponds to the frequency range of \gls{mbhb}~10.
Between ${3.02 \times 10^{-4}}$ and ${1.15 \times 10^{-2}}$\,Hz, we find that ${\Delta \sigma}$ has mean value of ${1.09 \times 10^{-2}}$\,rad.
This is below the pixel size ${\theta_{pix} = 1.6 \times 10^{-2}}$\,rad used in this analysis.
This shows that, in this region, coronagraphic \gls{tdi} is able to recover the sky location of the \gls{mbhb} up to pixel resolution, on average.

% Results - bias
However, this result is more nuanced if we look at the standard deviation associated with this mean value: ${1.69 \times 10^{-2}}$\,rad.
This means that, within the frequencies from this interval, the estimation can be off by more than a pixel even though the mean value is below $\theta_{pix}$.
This can be explained partially by the lower-frequency region of the interval, between ${3 \times 10^{-4}}$ and ${1 \times 10^{-3}}$\,Hz.
In this area, many frequencies display a ${\Delta \sigma}$ higher than $\theta_{pix}$.
This can be understood by the fact that the \gls{mbhb} signal is weaker in the lower-frequency region. 
Moreover, if we look, for example, at the frequencies of the simulated \glspl{vgb}, the largest bias observed is of $0.18$\,rad, which is about 10 pixels off.
This is consistent with the expectation from Sec.~\ref{sec:superposition-toy}, although a simulation with 49 \glspl{vgb} and an \gls{mbhb} is a less controlled setting where other parameters vary, not just the amplitude.

% Conclusion
%--------------------------------------
\subsection{Conclusion}

% Conclusion
The results presented in this section show how the problem of signal superposition manifests in the coronagraphic channel.
Through analyzing pairs of \glspl{gb}, we found that sources separated in frequency can be localized independently, even when they share the same sky position, while sources overlapping in frequency can only be localized if one of them is at least $\sim 100$ times louder than the other.
We extended this study to two types of sources (\glspl{gb} and \glspl{mbhb}) overlapping in frequency.
After analyzing a simulation of 49 \glspl{vgb} and a single \gls{mbhb}, we found that we were able to recover, on average, the sky location of the \gls{mbhb}, despite the presence of the \glspl{vgb}. 
However, for some frequencies, the coronagraphic \gls{tdi} estimate was biased.
For example, we identified that some \glspl{vgb} significantly biased the coronagraphic \gls{tdi} estimate by about 10 pixels.

% Ouverture
This limitation deserves to be further investigated in the future with more realistic simulations.
A population of ${\sim 10^7}$ \glspl{gb} might bias significantly the recovered position of a source of interest such as an \gls{mbhb}.
One way to bypass the signal superposition problem is to subtract known sources from the data~\cite{PhysRevD.107.063004,PhysRevD.111.103014}.
However, the effect of a population of \glspl{gb} on $\tilde{\kappa}$ has to be quantified before determining the best strategy to address this problem.
Such a study could not be carried out with the latest Sangria dataset, which also includes instrumental noise, because it only provides the Michelson variables, while coronagraphic \gls{tdi} requires Sagnac combinations.
This should be achievable with future common datasets which provide the necessary elements to compute \textit{any} \gls{tdi} variable~\cite{lisa_ddpc_mojito_light}.

%--------------------------------------
% Analysis in the presence of instrumental noise
%--------------------------------------
\section{Analysis in the presence of instrumental noise} \label{sec:analysis}

In this section, we address the problem of instrumental noise.
In this context, we define a statistical test based on the analytic noise model of coronagraphic \gls{tdi}.
We use this test to analyze a set of \gls{gb} simulations with instrumental noise and assess the applicability of the method.
Finally, we discuss implications and limitations of this work and future research directions.

% Chi2
%--------------------------------------
\subsection{Method} \label{sec:method}

% Data model
In the presence of noise, we model a data stream $\tilde{d}(f)$ as the superposition of noise $\tilde{n}(f)$ and signal $\tilde{s}(f)$, such that ${\tilde{d}(f) = \tilde{n}(f) + \tilde{s}(f)}$.
We also assume that when analyzing the three Sagnac channels in the frequency domain the noise can be described by the noise covariance matrix $S(f)$ (see \cref{eq:cov}).
Because coronagraphic \gls{tdi} is a linear combination of ${(\tilde{\alpha}, \tilde{\beta}, \tilde{\gamma})}$, the separation between signal and noise also holds for $\tilde{\kappa}$, such that
\begin{equation}
    \tilde{\kappa}(f, \delta, \alpha) = \tilde{\kappa}_n(f, \delta, \alpha) + \tilde{\kappa}_s(f, \delta, \alpha), \label{eq:model}
\end{equation}
where the $n$ subscript denotes the noise contribution and the $s$ subscript the signal contribution.

% Implication
Recalling the coronagraphic property, this data model implies that, as ${(f, \delta, \alpha) \rightarrow (f_\star, \delta_\star, \alpha_\star)}$, $\tilde{\kappa}_s$ tends to zero and ${\tilde{\kappa} \rightarrow \tilde{\kappa}_n}$.
As a consequence, in the presence of instrumental noise, the problem of finding a zero in $\tilde{\kappa}$ turns into the problem of finding \textit{just} noise in the data.
If we let $H_1$ be the signal hypothesis and $H_0$ be the null hypothesis, we can phrase this as: \textit{we consider there is a \gls{gw} source at sky position ${(\delta, \alpha)}$ if the observed data agrees with $H_0$}.

% Statistics
To be able to make this statement quantitative, we define
\begin{equation}
    z(\delta, \alpha) = \sum_{i} 4 \frac{|\tilde{\kappa}(f_i, \delta, \alpha)|^2 \times dt^2}{S_{\kappa\kappa}(f_i, \delta, \alpha)} df \label{eq:z}
\end{equation}
which compares, at frequencies ${f_i \in [f_{min}, f_{max}]}$, the null stream $\tilde{\kappa}$ to its analytic noise model $S_{\kappa\kappa}$.
In \cref{eq:z}, $dt$ is the inverse of the sampling frequency and ${df = 1/T}$ is the frequency resolution, $T$ being the duration of the analyzed time segment.
If there is only noise in the data and the noise model is well calibrated (see Appendix~\ref{sec:calibration}), one can show that ${z(\delta, \alpha) \sim \chi^2_{2N_f}}$~\cite{PhysRevD.63.042003}, where $N_f$ is the number of frequencies in ${[f_{min}, f_{max}]}$.
This means that we know the distribution of $z(\delta, \alpha)$ under $H_0$ and that anything which deviates significantly from this distribution indicates the presence of signal in the data.
This can be quantified by the probability of observing, under $H_0$, a value of $z(\delta, \alpha)$ greater than or equal to the observed value $z_{obs}(\delta, \alpha)$.
That is by computing the $p$ value
\begin{equation}
    p(\delta, \alpha) = P(z(\delta, \alpha) \geq z_{obs}(\delta, \alpha) | H_0),
\end{equation}
which can be calculated from the $\chi^2_{2N_f}$ distribution of $z$ under $H_0$.
When computing a $p$ value, one usually decides on a threshold $\beta$ (not to be confused with the Sagnac combination) \textit{below} which $H_0$ is \textit{rejected}.
Here, we instead use $\beta$ as a threshold \textit{above} which we \textit{do not reject} $H_0$.
This allows us to translate our previous statement into statistical terms: \textit{we consider there is a \gls{gw} source at sky position ${(\delta, \alpha)}$ if ${p(\delta, \alpha) > \beta}$}.

% Confidence interval
By applying this test across the sky, we define a region $\hat{\Omega}_\beta$ satisfying ${p(\delta, \alpha) > \beta}$.
This region corresponds to a \textit{confidence region}.
When only noise is present, $p(\delta, \alpha)$ is uniformly distributed~\cite{Cowan} (see also Appendix~\ref{sec:calibration}) and therefore ${p(\delta, \alpha) > \beta}$ with probability $(1 - \beta)$.
This is the case at the position of the source, where coronagraphic \gls{tdi} suppresses the signal and $\tilde{\kappa} \to \tilde{\kappa}_n$.
This implies that ${(\delta_\star, \alpha_\star) \in \hat{\Omega}_\beta}$ with probability ${(1 - \beta)}$, such that $\hat{\Omega}_\beta$ is a $(1 - \beta)$ confidence region.
In this section, we no longer provide point estimates of the position of the source, as in previous sections.
Instead, we provide an area $\hat{\Omega}_\beta$ which, across noise realizations, contains the true position of the source with probability $(1 - \beta)$.

% Summary
Based on this, we can outline the method we use to analyze coronagraphic \gls{tdi} in the presence of noise.
It follows four steps:
\begin{enumerate}
    \item Compute $\tilde{\kappa}(f,\delta,\alpha)$ and corresponding $S_{\kappa\kappa}(f, \delta, \alpha)$ across the sky on a frequency band $[f_{min}, f_{max}]$.
    \item Compute the statistic ${z(\delta, \alpha)}$.
    \item Compute the $p$ value associated with ${z(\delta, \alpha)}$.
    \item Determine the area $\hat{\Omega}_\beta$ which satisfies $p(\delta, \alpha) >\beta$.
\end{enumerate}
A few technical details are worth mentioning.
As concluded in Sec.~\ref{sec:tdi2}, because we include instrumental noise, we use \textit{second-generation} coronagraphic \gls{tdi} (and corresponding noise model) in our analysis.
In addition, in principle, the sky grid on which we perform the analysis is defined by intervals ${\delta \in [-\pi/2, \pi/2]}$ and ${\alpha \in [0, 2\pi]}$ sampled with $N_\delta$ and $N_\alpha$ points respectively.
In practice, we take advantage of the symmetry along the ${\delta=0}$\,rad plane in the \gls{lisa} frame and only do the computations on the northern hemisphere in the same frame.
This symmetry also manifests in $\hat{\Omega}_\beta$, which accounts for the $p(\delta, \alpha) >\beta$ regions on both hemispheres (see \cref{fig:p-1000}).
Because coronagraphic \gls{tdi} cannot distinguish between the true position of the source and its mirror position, we report $\hat{\Omega}_\beta$ as the total area of both regions, which is the area one would effectively have to consider.

% Parameters
In this analysis, three parameters can be tuned:
\begin{itemize}
    \item The frequency window ${[f_{min}, f_{max}]}$ over which $z$ is computed.
    \item The number of pixels in the sky map or, equivalently, the pixel size $\theta_{pix}$.
    This can also be expressed as a different number of pixels $N_{pix}$ used to sample a rectangular grid (i.e., ${N_\delta=N_\alpha=N_{pix}}$).
    \item The threshold $\beta$, i.e., the selection criterion.
\end{itemize}
In the results presented in the next section, we consider different choices of ${[f_{min}, f_{max}]}$ and $\beta$, while keeping $N_{pix}$ fixed.
This is because we found that the value of ${N_{pix} = 60}$ (corresponding to a pixel size at the equator of $19\,\mathrm{deg}^2$) achieved a sufficient resolution for our studies, as we will discuss later.

% Table: GB params
\begin{table}[tbp]
\begin{ruledtabular}
\begin{tabular}{c c c}
Parameter & \gls{gb} & Units \\
\colrule
$\mathcal{A}$ & $\{1.22, 2.44, 3.67, 6.11, 8.56, 12.2\} \times 10^{-20}$ & - \\
$f_0$ & 3.79 & mHz \\
$\dot{f}$ & $1.14 \times 10^{-16}$ & Hz/s \\
$\iota$ & 1.43 & rad \\
$\phi$ & 0.0 & rad \\
$\Psi$ & 0.0 & rad \\
$\delta_\star$ & 0.92 & rad \\
$\alpha_\star$ & 5.95 & rad \\
\end{tabular}
\end{ruledtabular}
\caption{\label{tab:sim-params-gbs}Simulation parameters of the six injections used to assess coronagraphic \gls{tdi}'s ability to localize sources in the presence of instrumental noise.  Parameters $\{f_0, \dot{f}, \iota, \phi, \Psi\}$ could be arbitrary but were taken to be those of a \gls{vgb}: {ZTFJ2243}\,\cite{kupfer2023lisa}. The parameters $\{\Psi, \delta_\star, \alpha_\star\}$ are given in the \gls{icrs} frame, following \textsc{LISA GW Response}'s convention.
The data are simulated at a sampling rate of ${0.2}$\,Hz for 5\,days ($8.64 \times 10^4$ data points).
We choose a lower sampling rate than the realistic 4\,Hz to reduce computational cost.}
\end{table}
    
% Applicability
%--------------------------------------
\subsection{Applicability} \label{sec:applicability}

% Simulation
To assess the applicability of this method, we create a set of simulations of six individual \glspl{gb} (see \cref{tab:sim-params-gbs}) with instrumental noise.
We run 1\,000 noise simulations with parameters given in \cref{tab:noise-sims} under simulation D.
For a single \gls{gb}, we then obtain 1\,000 simulations with different noise realizations by adding the \gls{gb} signal to the  1\,000 noise simulations.
This way, we obtain a set of 6\,000 simulations from the same underlying noise realizations.
The amplitudes of the six \glspl{gb} were chosen such that a range of \glspl{snr} is explored.
The \gls{snr} we quote here is calculated from the Sagnac variables ${(\tilde{\alpha}_2, \tilde{\beta}_2, \tilde{\gamma}_2)}$ and the covariance matrix $S(f)$ given in \cref{eq:cov} in Sec.~\ref{sec:tdi2-noise}.

% Table: violin stats
\begin{table}[tbp]
\begin{ruledtabular}
\begin{tabular}{c c c c c}
$\mathcal{A}$ & Median & Null rate & Miss rate & Median \\
 & \gls{snr} & & & $\hat{\Omega}_{0.1}$ [$\mathrm{deg}^2$] \\
\hline
$1.22 \times 10^{-20}$ & 1\,243 & 0.09 & 0.10 & $4.13\times 10^{4}$ \\
$2.44 \times 10^{-20}$ & 2\,486 & 0.09 & 0.10 & $2.6\times10^{4}$ \\
$3.67 \times 10^{-20}$ & 3\,729 & 0.09 & 0.10 & $2.0\times10^{4}$ \\
$6.11 \times 10^{-20}$ & 6\,215 & 0.09 & 0.10 & $1.1\times10^{4}$ \\
$8.56 \times 10^{-20}$ & 8\,701 & 0.09 & 0.10 & $6.2\times10^{3}$ \\
$12.2 \times 10^{-20}$ & 12\,430 & 0.09 & 0.10 & $3.4\times10^{3}$ \\
\end{tabular}
\end{ruledtabular}
\caption{\label{tab:violin-stats}Descriptive statistics associated with \cref{fig:violin}. We report the fraction of noise realizations for which ${\hat{\Omega}_{0.1} = 0}$ (null rate) and for which ${(\delta_\star, \alpha_\star) \notin \hat{\Omega}_{0.1}}$ (miss rate). We also provide the median $\hat{\Omega}_{0.1}$ obtained for the six injections (see \cref{tab:sim-params-gbs}) over 1\,000 noise realizations.}
\end{table}

% Reference analysis
%--------------------------------------
\subsubsection{Reference analysis} \label{sec:reference}

% Parameters
We first apply the method with a set of conservative parameters, which we use as a reference in the rest of this section.
We begin with a frequency window ${[3.56, 4.02]}$\,mHz centered around the characteristic frequency ${f_0=3.79}$\,mHz of the injected \glspl{gb}.
Given the short duration of the simulation and the frequency derivative of the simulated \glspl{gb}, the expected drift from $f_0$ in this simulation is of the order of $10^{-11}$\,Hz, orders of magnitude below the frequency resolution of the simulation (${df = 2.33\,\mu}$Hz).
Lastly, we use the value of ${\beta = 0.1}$, which implies that ${\hat{\Omega}_{0.1}}$ should fail to contain the true position of the source only in 10\,\% of the noise realizations.

% Figure: violin plot
\begin{figure}[tbp]
    \centering
    \includegraphics[width=\linewidth]{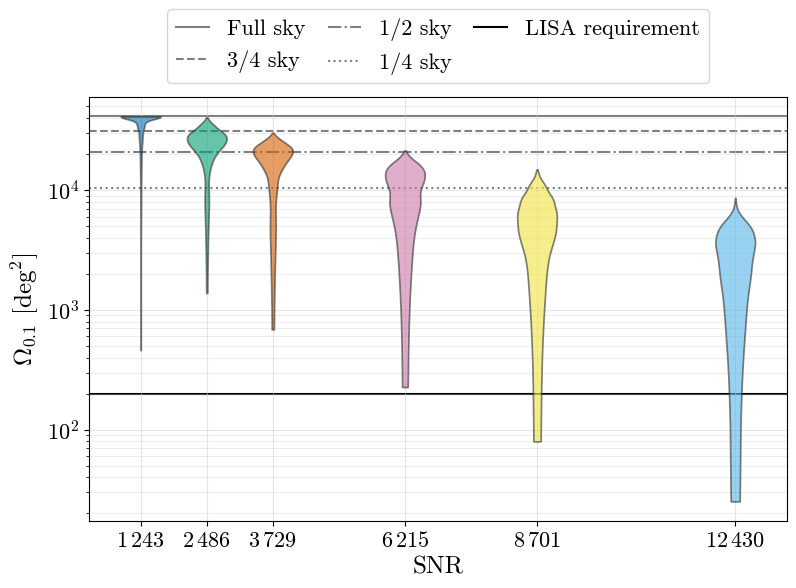}
    \caption{\label{fig:violin}90\% confidence regions for the six injections whose parameters are given in \cref{tab:sim-params-gbs}. The distributions show results across 1\,000 noise realizations, excluding cases where ${\hat{\Omega}_{0.1} = 0}$. Excluded realizations represent {0.09} of the total realizations, as reported in \cref{tab:violin-stats}.}
\end{figure}

% Results
The results are shown in \cref{fig:violin} and \cref{tab:violin-stats}.
Although $\hat{\Omega}_{0.1}$ decreases with \gls{snr}, it remains large across the range of \glspl{snr} considered.
The simulated sources range from \gls{snr} $1\,243$ to $12\,430$, far above what \glspl{gb} reach over a few days but from typical to beyond the expected \gls{snr} of \gls{lisa} \glspl{mbhb}~\cite{colpi_lisa_2024}.
This range allows us to assess the performance of coronagraphic \gls{tdi} in favorable conditions.
Even in the most favorable scenario of \gls{snr} $12\,430$, \cref{tab:violin-stats} reports a median $\hat{\Omega}_{0.1}$ across noise realizations of ${3.4 \times 10^3\,\mathrm{deg}^2}$.
For comparison, although the two quantities are not strictly equivalent and the sources considered are different, \gls{lisa} requirements ask for a sky-localization precision of less than $200\,\mathrm{deg}^2$ 4\,hours before merger~\cite{colpi_lisa_2024}.
This requirement is not met by the sky-localization method proposed here.
For completeness, the median quoted here is computed across noise realizations, excluding cases where ${\hat{\Omega}_{0.1} = 0}$ (reported in \cref{tab:violin-stats}).
Such cases correspond to ${p(\delta, \alpha) < \beta}$ across the sky and imply that ${(\delta_\star, \alpha_\star) \notin \hat{\Omega}_{0.1}}$.
It follows that the fraction of ${\hat{\Omega}_{0.1} = 0}$ cases (null rate) is a lower bound on the fraction of noise realizations in which $\hat{\Omega}_{0.1}$ misses the source (miss rate).
The values reported in \cref{tab:violin-stats} are consistent with this argument.
We found a miss rate of $0.10$, as expected for ${\beta = 0.10}$, with a lower null rate of $0.09$.
The difference between the two rates corresponds to cases where some pixels satisfied ${p(\delta, \alpha) > \beta}$, but the selected region missed the true position of the source, i.e., ${(\delta_\star, \alpha_\star) \notin \hat{\Omega}_\beta}$.

\begin{figure*}[hp]
    \centering
    % Row 1
    \begin{subfigure}[t]{0.48\textwidth}
        \centering
        \includegraphics[width=\linewidth,height=0.19\textheight,keepaspectratio]{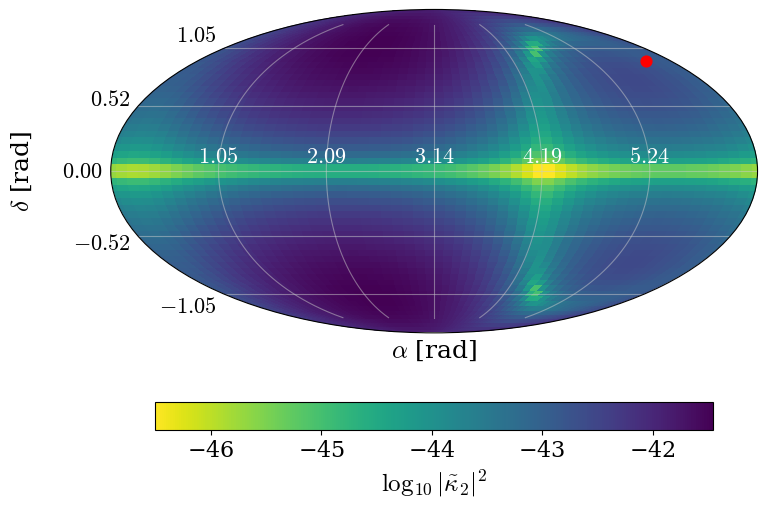}
        \caption{\label{fig:kappa-100}Quietest injection. Second-generation coronagraphic \gls{tdi} sky map at the characteristic frequency of the source ${f_0 = 3.79}$\,mHz.}
    \end{subfigure}
    \hfill
    \begin{subfigure}[t]{0.48\textwidth}
        \centering
        \includegraphics[width=\linewidth,height=0.19\textheight,keepaspectratio]{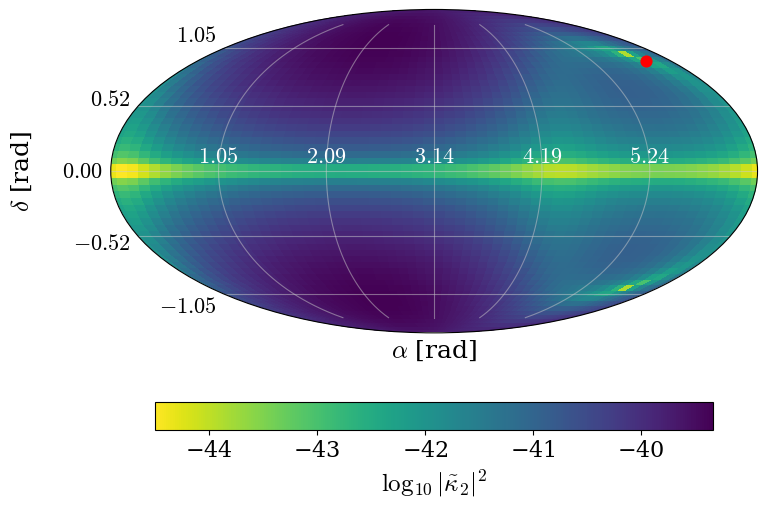}
        \caption{\label{fig:kappa-1000}Loudest injection. Second-generation coronagraphic \gls{tdi} sky map at the characteristic frequency of the source ${f_0 = 3.79}$\,mHz.}
    \end{subfigure}

    \vspace{0.2cm}

    % Row 2
    \begin{subfigure}[t]{0.48\textwidth}
        \centering
        \includegraphics[width=\linewidth,height=0.19\textheight,keepaspectratio]{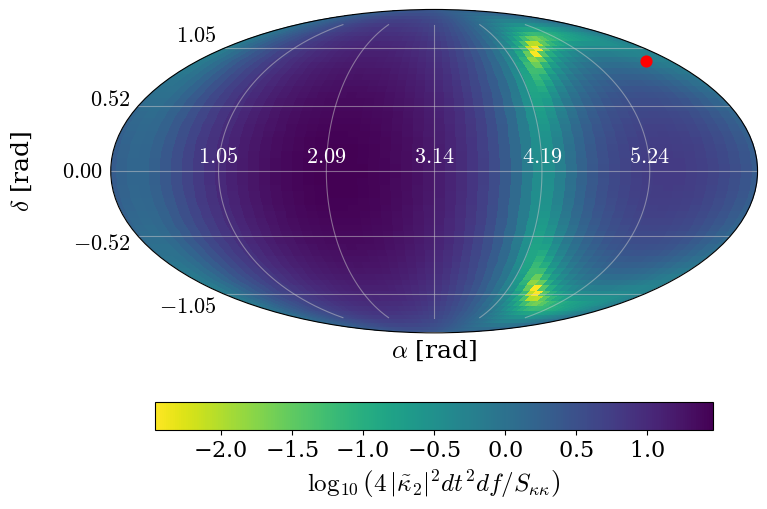}
        \caption{\label{fig:kappa-over-noise-100}Quietest injection. Second-generation coronagraphic \gls{tdi} normalized sky map at the characteristic frequency of the source ${f_0 = 3.79}$\,mHz.}
    \end{subfigure}
    \hfill
    \begin{subfigure}[t]{0.48\textwidth}
        \centering
        \includegraphics[width=\linewidth,height=0.19\textheight,keepaspectratio]{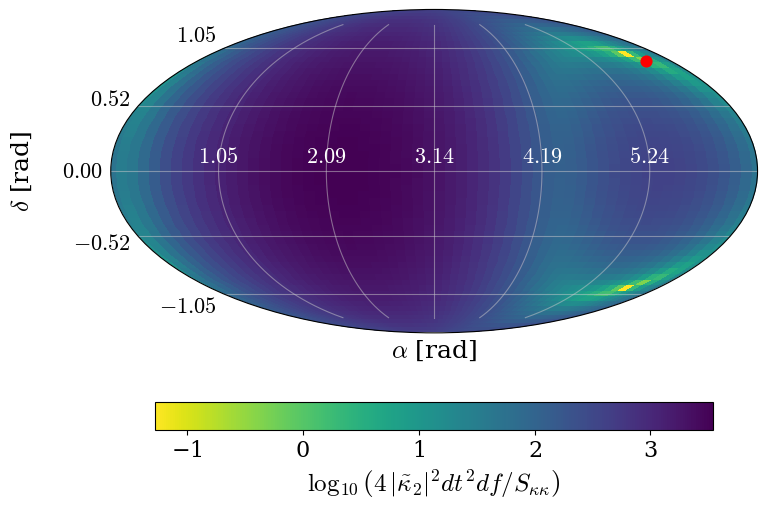}
        \caption{\label{fig:kappa-over-noise-1000}Loudest injection. Second-generation coronagraphic \gls{tdi} normalized sky map at the characteristic frequency of the source ${f_0 = 3.79}$\,mHz.}
    \end{subfigure}

    \vspace{0.2cm}

    % Row 3
    \begin{subfigure}[t]{0.48\textwidth}
        \centering
        \includegraphics[width=\linewidth,height=0.19\textheight,keepaspectratio]{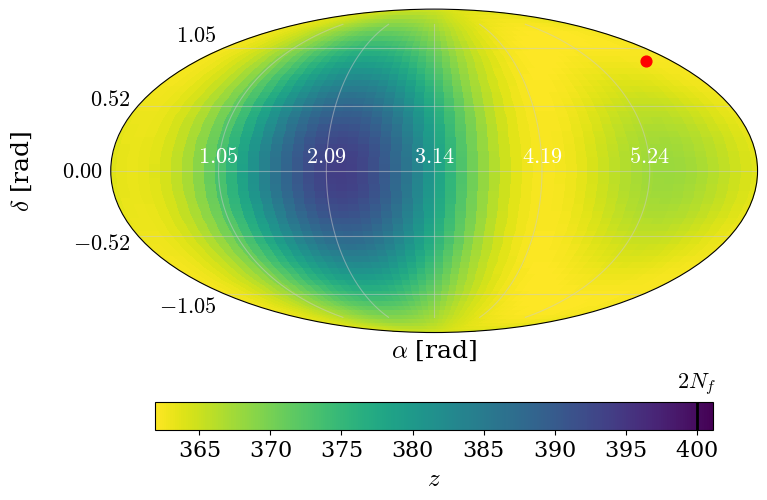}
        \caption{\label{fig:z-100}Quietest injection. $z$ statistic computed over $N_f=200$ frequency bins (${df = 2.33\,\mu}$Hz).}
    \end{subfigure}
    \hfill
    \begin{subfigure}[t]{0.48\textwidth}
        \centering
        \includegraphics[width=\linewidth,height=0.19\textheight,keepaspectratio]{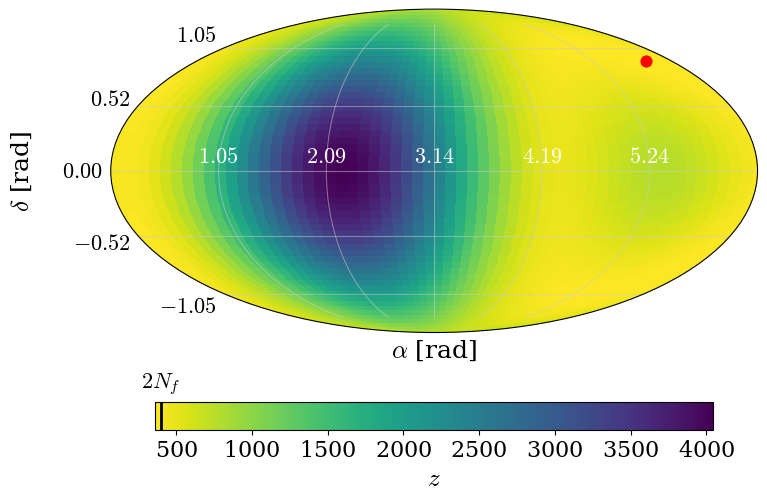}
        \caption{\label{fig:z-1000}Loudest injection. $z$ statistic computed over $N_f=200$ frequency bins (${df = 2.33\,\mu}$Hz).}
    \end{subfigure}

    \vspace{0.2cm}

    % Row 4
    \begin{subfigure}[t]{0.48\textwidth}
        \centering
        \includegraphics[width=\linewidth,height=0.19\textheight,keepaspectratio]{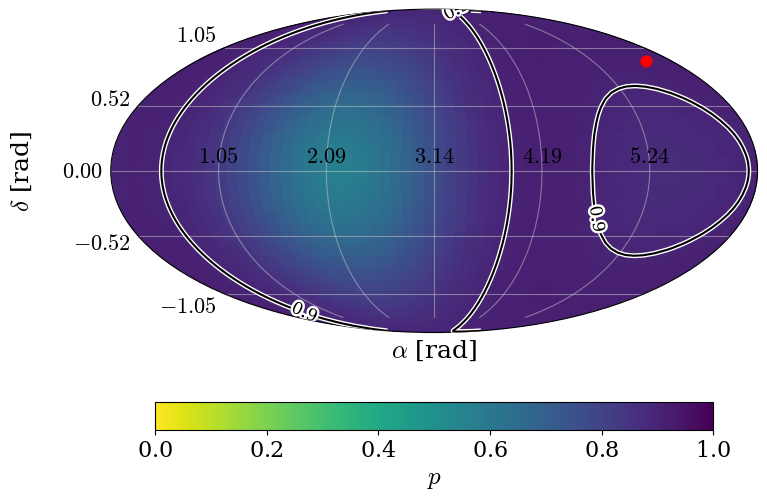}
        \caption{\label{fig:p-100}Quietest injection. $p$ value sky map derived from \cref{fig:z-100}.}
    \end{subfigure}
    \hfill
    \begin{subfigure}[t]{0.48\textwidth}
        \centering
        \includegraphics[width=\linewidth,height=0.19\textheight,keepaspectratio]{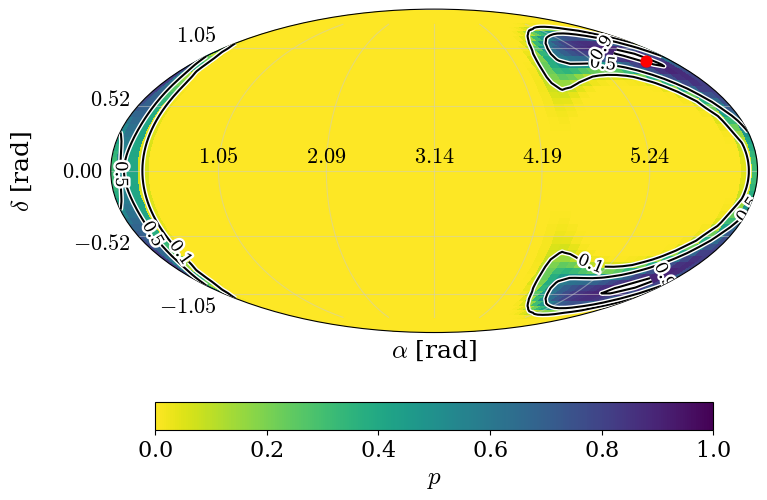}
        \caption{\label{fig:p-1000}Loudest injection. $p$ value sky map derived from \cref{fig:z-1000}.}
    \end{subfigure}

    \caption{\label{fig:sky-map-steps}Sky maps in the \gls{lisa} frame of the intermediate steps used in the analysis (see Sec.~\ref{sec:method}) of the loudest and quietest injections (see \cref{tab:sim-params-gbs}) for one noise realization (see simulation D in \cref{tab:noise-sims}). The left column corresponds to the quietest injection (\gls{snr} 1\,243), while the right column shows the loudest injection (\gls{snr} 12\,430) results. The red point represents the position of the injection in the \gls{lisa} frame.}
\end{figure*}

% Comment
To better understand this result, it is insightful to look at the output of each intermediate step described in Sec.~\ref{sec:method}.
This is shown for both the quietest and loudest injection for one particular noise realization in \cref{fig:sky-map-steps}.
Inspecting this figure, we realize there are two main effects behind the results in \cref{fig:violin} and \cref{tab:violin-stats}:
\begin{itemize}
    \item \textit{Dilution:} At the characteristic frequency of the source, i.e., in \cref{fig:kappa-100,fig:kappa-1000,fig:kappa-over-noise-100,fig:kappa-over-noise-1000}, we see two clear symmetric spots close to the true position of the source and its mirror.
    However, when we sum over all frequencies in ${[f_{min}, f_{max}]}$ to compute $z$, we include uninformative sky maps and \textit{dilute} the information.
    This is why the clear spots become much broader regions in \cref{fig:z-100,fig:z-1000}.
    \item \textit{Significance:} Although \cref{fig:z-100,fig:z-1000} look similar, the variations in \cref{fig:z-1000} are much more \textit{statistically significant} than those in \cref{fig:z-100}.
    Values of $z$ in \cref{fig:z-100} fluctuate around ${2N_f}$, as expected for noise (see Appendix~\ref{sec:calibration}).
    This is reflected in the $p$ value map \cref{fig:p-100}, where we see that the whole sky is well above the threshold ${\beta = 0.1}$.
\end{itemize}
The first effect can be reduced by narrowing the frequency window, while the threshold $\beta$ sets how strictly regions compatible with noise are selected.
Next, we investigate the impact of each of these parameters on the results presented in \cref{fig:violin}.

% Number of bins
%--------------------------------------
\subsubsection{Frequency window} \label{sec:freq-window}

% Set up
To study the impact of different frequency windows on the final $\hat{\Omega}_{0.1}$, we consider two frequency windows: ${[3.67,3.90]}$\,mHz and ${[3.77,3.81]}$\,mHz.
Both are shorter than the reference ${[3.56,4.02]}$\,mHz.
The first window reduces the number of frequency bins $N_f$ in the window by a factor 2, while the second goes down to a factor 10 (see \cref{tab:improvements}).
Both windows are centered around the characteristic frequency of the simulated \glspl{gb}.

% Table: improvements
\begin{table}[tbp]
\begin{ruledtabular}
\begin{tabular}{c c c c}
$N_f$ & $\beta$ & Null rate & $\hat{\Omega}_{ref}/\hat{\Omega}_\beta$ \\
\colrule
200 & 0.1 & 0.09 & 1 \\
100 & 0.1 & 0.09 & $1.21 \pm 0.11$\\
20 & 0.1 & 0.06 & $2.16 \pm 0.65$\\
\colrule
20 & 0.1 & 0.06 & 1 \\
20 & 0.5 & 0.38 & $2.13 \pm 0.39$\\
20 & 0.9 & 0.83 & $4.16 \pm	1.16$\\
\end{tabular}
\end{ruledtabular}
\caption{\label{tab:improvements}Comparison between the \textit{frequency window} and \textit{threshold} analyses (see \cref{fig:improvements}). Number of frequency bins $N_f$ and threshold $\beta$ used are shown with the fraction of realizations for which ${\hat{\Omega}_\beta = 0}$ (null rate). The ratio ${\hat{\Omega}_{ref}/\hat{\Omega}_\beta}$ represents the average, over all six injections, of the ratio between the median reference $\hat{\Omega}_{ref}$ and the median $\hat{\Omega}_\beta$ obtained with the specified $N_f$ and $\beta$ settings. For the first series, with $\beta$ fixed, the result for ${N_f=200}$ and ${\beta=0.1}$ is taken as a reference. For the second series, where ${N_f = 20}$ is fixed, the result for ${N_f=20}$ and ${\beta = 0.1}$ is the reference. The average ratio ${\hat{\Omega}_{ref}/\hat{\Omega}_\beta}$ is reported with the standard deviation across the six datasets.}
\end{table}

\begin{figure}[tbp]
    \centering
    % 1st generation
    \begin{subfigure}[t]{\columnwidth}
        \centering
        \includegraphics[width=0.96\linewidth]{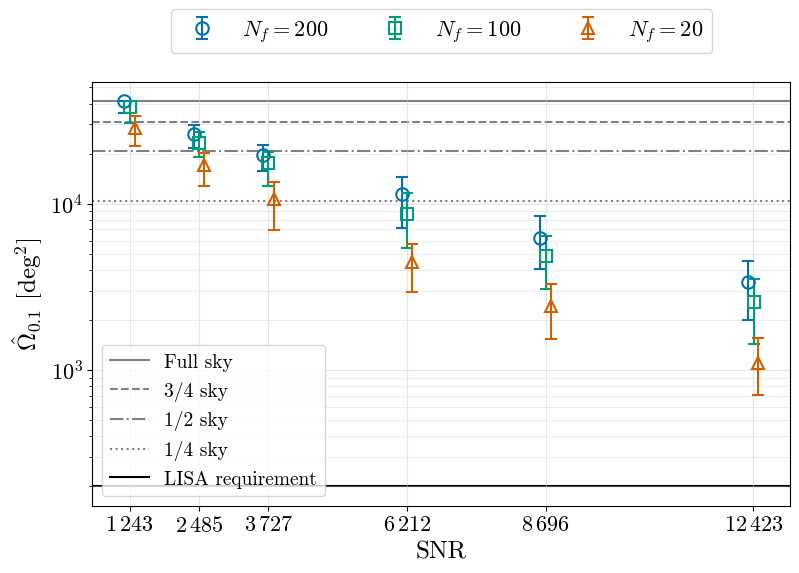}
        \caption{\label{fig:nf}Results obtained for different frequency windows centered around the characteristic frequency of the source ${f_0 = 3.79}$\,mHz  and fixed threshold ${\beta=0.1}$. The number of frequency bins ${N_f=200}$ corresponds to a ${[3.56,4.02]}$\,mHz window, ${N_f=100}$ to ${[3.67,3.90]}$\,mHz and ${N_f=20}$ to ${[3.77,3.81]}$\,mHz (${df = 2.33\,\mu\mathrm{Hz}}$ is fixed).}
    \end{subfigure}
    \vspace{0.3cm}
    % 2nd generation
    \begin{subfigure}[t]{\columnwidth}
        \centering
        \includegraphics[width=0.96\linewidth]{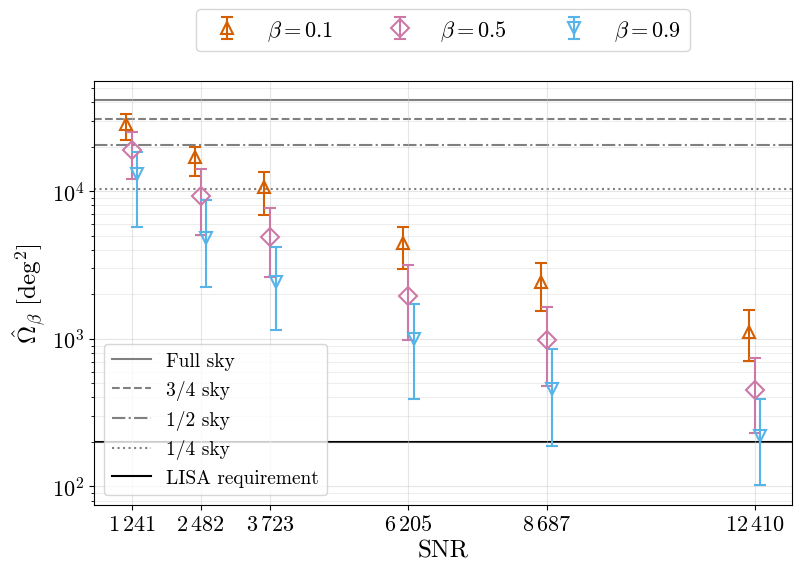}
        \caption{\label{fig:beta}Results obtained for different threshold values $\beta$ and fixed frequency window ${[3.67,3.90]}$\,mHz centered around the characteristic frequency ${f_0 = 3.79}$\,mHz of the injections (${N_f = 20}$ and ${df = 2.33\,\mu\mathrm{Hz}}$).}
    \end{subfigure}
    \caption{\label{fig:improvements}Confidence regions for the six injections (see \cref{tab:sim-params-gbs}) obtained with different analysis settings $N_f$ and $\beta$. Points represent the median $\hat{\Omega}_\beta$ computed over all noise realizations, excluding zero-area results (see \cref{tab:improvements} for the fraction of excluded realizations). The error bars correspond to the first and third quartile. The points are slightly separated around \gls{snr} for visibility, points obtained for the same injection have the same \gls{snr}. Lines show the fraction of sky a given $\hat{\Omega}_\beta$ represents.}
\end{figure}

% Results
Results are presented in \cref{fig:nf}.
The blue circles in \cref{fig:nf} correspond to the median values of the distributions in \cref{fig:violin} and the error bars represent the first and third quartiles of the same distributions.
In this figure, we summarize the results for $N_f \in \{200, 100, 20\}$. 
We find that $\hat{\Omega}_{0.1}$ decreases with $N_f$, thus confirming the dilution effect mentioned in the reference analysis.
For instance, the median $\hat{\Omega}_{0.1}$ for the quietest source (\gls{snr} 1\,243) goes down from the full sky to less than half of the sky by reducing the number of frequency bins used by a factor of 10.
If we do the comparison across all six injections, we find an improvement of a factor of ${2.16}$ on average (with a standard deviation of $0.65$) between the reference result and the analysis with ${N_f=20}$ (see \cref{tab:improvements}).
This result shows that narrowing down the interval of interest ${[f_{min}, f_{max}]}$ before performing a coronagraphic-\gls{tdi}-based analysis offers a clear advantage.
One has to maximize the number of \textit{informative} frequencies in ${[f_{min}, f_{max}]}$.
It is therefore important to make an informed decision on the frequency window to be analyzed, potentially based on other methods.

% Figure: improvements

% Threshold
%--------------------------------------
\subsubsection{Threshold} \label{sec:threshold}

% Shrinkage
To investigate the impact of the threshold $\beta$ on the analysis, we consider three values $\beta \in \{0.1, 0.5, 0.9\}$, while keeping ${N_f = 20}$ (according to Sec.~\ref{sec:freq-window}) fixed.
The results are presented in \cref{fig:beta}.
Similar to \cref{fig:nf}, only the quartiles of the $\hat{\Omega}_\beta$ distributions are reported.
As one could expect from \cref{fig:p-1000}, the $\hat{\Omega}_\beta$ decrease with higher threshold values.
A higher $\beta$ is a stricter condition for a pixel to be kept in $\hat{\Omega}_\beta$: only pixels with $p(\delta, \alpha) > \beta$ pass the test.
As a consequence, fewer pixels pass and $\hat{\Omega}_\beta$ shrinks.
Indeed, \cref{tab:improvements} reports a median $\hat{\Omega}_{0.9}$ smaller than $\hat{\Omega}_{0.1}$ by a factor of ${4.16 \pm 1.16}$ for the same $N_f$.

% Limitation
Looking at the values of $\hat{\Omega}_\beta$ alone, there is a clear advantage to use a high threshold.
However, this improvement comes at the cost of a higher miss rate.
With ${\beta = 0.9}$, {83\,\%} of the noise realizations had ${\hat{\Omega}_{0.9} = 0}$ (see \cref{tab:improvements}).
This comes back to the fact that at the position of the source, ${\tilde{\kappa} \rightarrow \tilde{\kappa}_n}$ and $p$ is uniformly distributed.
As a consequence, even if the \gls{gw} signal is correctly suppressed, $p$ is more likely to be above $0.1$ than above $0.9$.
It follows that with a threshold at $0.9$, we are more likely to fail to satisfy the condition ${p > 0.9}$ and therefore miss the source.
This is also consistent with the interpretation of $\hat{\Omega}_\beta$ as a ${(1 - \beta)}$ confidence region.
If ${\beta = 0.9}$, then ${(\delta_\star, \alpha_\star)}$ should be in $\hat{\Omega}_{0.9}$ only in {10\,\%} of the noise realizations, so with a miss rate of {$90$\,\%}.
Again, the null rate ({$83$\,\%}) is consistent with but not exactly {$90$\,\%} because it only provides a lower bound for miss rate, as already discussed.

% Conclusion
%--------------------------------------
\subsubsection{Conclusion}

% Summary
In this section, we applied a statistical test based on coronagraphic \gls{tdi} to estimate the ${(1 - \beta)}$ confidence region $\hat{\Omega}_\beta$ of six synthetic \glspl{gb} in the presence of instrumental noise, spanning a wide range of \glspl{snr}.
We first analyzed the six datasets with conservative parameters ${N_f = 200}$ and ${\beta = 0.1}$.
With these settings, the median $\hat{\Omega}_{0.1}$ of the loudest source (\gls{snr} $12\,430$) is ${3.4 \times 10^3\,\mathrm{deg}^2}$, i.e., about $8\,\%$ of the sky (see \cref{tab:violin-stats}).
We then showed that narrowing the frequency window around the characteristic frequency of the injection to ${N_f = 20}$ reduces $\hat{\Omega}_{0.1}$ by a factor $2.16 \pm 0.65$ on average, at the same confidence level (see \cref{tab:improvements}).
Finally, increasing the threshold to $\beta = 0.9$ reduces the median area by a further factor $4.16 \pm 1.16$, but at the cost of a much higher miss rate (from $\sim 10\,\%$ to $\sim 90\,\%$), since $\hat{\Omega}_{0.9}$ is only a 10\,\% confidence region.
This manifests as higher null rates in \cref{tab:improvements}, although null rates are lower than expected miss rates as already explained.

% Implications and limitations
The main limitation highlighted in this section is that, without any assumption on the nature of the source, constraining its sky location requires very high \gls{snr}.
Our best results were achieved for \gls{snr} $12\,430$, while injections with \gls{snr} ${\sim 10^3}$, typical for \gls{lisa} \glspl{mbhb}~\cite{colpi_lisa_2024}, were poorly constrained.
For instance, for the injection with \gls{snr} $2\,486$ and ${N_f = 20}$, the median $\hat{\Omega}_{0.1}$ is just below half of the sky (see \cref{fig:nf}).
For this reason, it is unlikely that coronagraphic \gls{tdi}, in its current model-agnostic form, could provide a useful prior for parameter estimation, although this could be explored in the future.
This result also challenges its application to model-agnostic low-latency searches, given that much lower \glspl{snr} are expected during pre-merger.
For example, most systems considered in a study on the multi-messenger perspectives of joint \gls{lisa} and \gls{athena} detections did not reach \gls{snr} ${\sim 10^3}$ even one hour before merger~\cite{piro_chasing_2023}.
A second limitation is the need for prior knowledge of the frequency window ${[f_{\min}, f_{\max}]}$ containing the signal, which would have to be obtained with another method.
Next, we discuss these results and propose future research directions.

% Discussion
%--------------------------------------
\subsection{Discussion} \label{sec:discussion}

% Previous (and other) results
The results presented in this paper represent a significant step towards the application of coronagraphic \gls{tdi} to realistic \gls{lisa} data.
In our previous paper, only a few scenarios including instrumental noise were considered and no statistical test based on coronagraphic \gls{tdi} was developed.
The results there, in particular Fig.\,11 which is analogous to \cref{fig:kappa-100,fig:kappa-1000} in this paper, pointed to a similar conclusion to the one reached here: coronagraphic \gls{tdi} alone can only be directly applied to sufficiently loud sources.
This qualitative statement has been confirmed quantitatively by the more thorough analysis presented here.
As a consequence, the definitive result to retain is the one presented here: this approach cannot meaningfully constrain the sky location of \gls{gw} sources of less than a few thousand \gls{snr}.

% Model
This is a constraint on how much information one can extract from the response of \gls{lisa} to \glspl{gw} alone.
In principle, this can be improved by introducing some modeling of the source.
Intuitively, this corresponds to extracting information from \cref{fig:kappa-100,fig:kappa-1000} not just by comparing it against a noise model, but also a model for the \gls{gw} signal.
Specifically, this means modeling the residual ${\tilde{\kappa}_s(f, \delta, \alpha)}$ term which is not suppressed at ${(f, \delta, \alpha) \neq (f_\star, \delta_\star, \alpha_\star)}$.
This has been done for null-stream channels in the context of \glspl{pta} in~\cite{10.1093/mnras/sty892}, where they derive a likelihood function specific to these channels.
It would be interesting to apply their methodology to coronagraphic \gls{tdi}.

% Time-freq
Another research direction, already mentioned in Sec.~\ref{sec:tdi2-conclusion}, is generalizing coronagraphic \gls{tdi} to the time-frequency domain.
Different approaches can be investigated.
One is to derive time-varying coronagraphic coefficients based on the time-frequency response of \gls{lisa} based on~\cite{marsat_exploring_2021}.
Another option is to do it piecewise, assuming \gls{lisa} to be frozen on short time segments and to update coefficients ${\vec{A}(f, \delta, \alpha)}$ accordingly, similar to what was done in~\cite{costa_barroso_coronagraphic_2025}.
In both cases, this should allow for longer observation windows, therefore leading to more \gls{snr} accumulated over time.
This is crucial for a coronagraphic-\gls{tdi}-based method which requires high \gls{snr} to recover the sky location of a \gls{gw} source, such as the one presented in Sec.~\ref{sec:analysis}.
Moreover, as demonstrated in~\cite{marsat_exploring_2021}, introducing time dependence in the response breaks the North and South degeneracy observed in the \gls{lisa} frame (see \cref{fig:sky,fig:sky-map-steps}).
Breaking this degeneracy could divide the $\hat{\Omega}_\beta$ by up to a factor of 2.

% Glitches
A potential application of coronagraphic \gls{tdi}, which has not been considered in this paper but already suggested in~\cite{costa_barroso_coronagraphic_2025}, is as a \gls{gw} discriminator.
This is based on the fact that coronagraphic \gls{tdi} extracts information from the coherent geometric response of \gls{lisa} to \glspl{gw}.
As a consequence, it responds differently for \glspl{gw} and instrumental glitches which impact the constellation differently.
In the context of ground-based detectors, such null-stream vetoes have been suggested in~\cite{chatterji_coherent_2006,wen_coherent_2005,ajith_null-stream_2006,harry_targeted_2011}.
A particularly interesting extension of this work would be to test the discriminating statistics proposed in~\cite{chatterji_coherent_2006}, which could be easily implemented in \textsc{NullStream}.

%--------------------------------------
% Conclusion
%--------------------------------------
\section{Conclusion}

% Context
In this paper, we have addressed three challenges which arise in realistic \gls{lisa} data: \textit{(i)} spacecraft motion; \textit{(ii)} signal superposition; \textit{(iii)} instrumental noise.
This represents an important step towards the concrete application of coronagraphic \gls{tdi} to \gls{lisa} data-analysis problems.

% Section (ii)
First, we have extended coronagraphic \gls{tdi} to second generation by replacing the first-generation Sagnac variables with their second-generation equivalent, while keeping the coronagraphic coefficients unchanged.
We have derived the corresponding analytic noise model and validated it against instrument simulations.
In the presence of spacecraft motion and laser frequency noise, $\tilde{\kappa}_2$ reduces laser frequency noise down to secondary noise levels, whereas the first-generation combination exceeds them by nearly three orders of magnitude.
Moreover, $\tilde{\kappa}_2$ preserves the coronagraphic property: its minimum lies at $3$\,deg from the position of the source, against $44$\,deg for $\tilde{\kappa}$.

% Section (iii)
Second, we have shown that coronagraphic \gls{tdi} can localize different sources present in the data, provided they do not overlap in frequency, even when they share the same sky position.
When two sources overlap in frequency, only the dominant one can be localized, and only if it is at least ${\sim 100}$ times louder.
These conclusions are consistent with the analysis of a more complex simulation containing an \gls{mbhb} and 49 \glspl{vgb}.
In this case, the sky location of the \gls{mbhb} is recovered within one pixel on average across its frequency band, with larger biases at the low-frequency end of the band, where the \gls{mbhb} signal is weaker, and at the characteristic frequencies of some \glspl{vgb}.

% Section (iv)
Third, we have introduced a statistical test comparing coronagraphic \gls{tdi} to its analytic noise model, which defines a ${(1-\beta)}$ confidence region $\hat{\Omega}_\beta$ for the sky position of the source.
We have verified that these regions have the expected coverage.
However, applying this test to \glspl{gb} with \glspl{snr} between $10^3$ and $10^4$, we found that $\hat{\Omega}_{0.1}$ remains large, with a median of ${3.4 \times 10^3}\,\mathrm{deg}^2$ (about 8\,\% of the sky) at the highest \gls{snr}.
Narrowing the frequency window around the source improves this result by a factor ${\sim 2}$ at the same confidence level, which is still well above the \gls{lisa} requirement of $200\,\mathrm{deg}^2$ for low-latency searches.

% Ouverture
Together, these results set a limit on the information that can be extracted from the geometric response of \gls{lisa} alone.
Coronagraphic \gls{tdi} is robust to realistic instrumental conditions and to signal superposition, but on its own it only constrains the sky location of high-\gls{snr} sources, proving impractical for low-latency searches.
Improving on this limit requires partially modeling the signal, for instance through a likelihood built on coronagraphic \gls{tdi}, and longer observation windows, enabled by time-dependent coronagraphic coefficients.
Independently of sky localization, the information from the coherent geometric response of \gls{lisa} captured by coronagraphic \gls{tdi} makes it a promising tool to discriminate \gls{gw} signals from instrumental glitches.
The \textsc{NullStream} package, which implements first- and second-generation coronagraphic \gls{tdi} and corresponding noise models, should facilitate these future investigations.

%--------------------------------------
% Acknowledgements
%--------------------------------------
\begin{acknowledgments}
% People
The authors thank François Mauger, Benoît Revenu, Jean-Baptiste Bayle, Aurélien Hees, Martin Staab, Dam Quang Nam, Quentin Baghi and Michael J. Williams for insightful discussions.
The authors also thank the \gls{lisa} Simulation Expert Group together with the \gls{lisa} Data Challenge Working Group for developing and supporting the simulation tools used in this work.
% Cimputing
This work also used computing resources from the {CC-IN2P3} (Lyon, France).
% Funding
RCB and IH acknowledge support from the \gls{uksa}, through grant UKRI968, supporting the UK's contribution to \gls{lisa}'s Ground Segment activities.
% Software: numpy, scipy, matplotlib
This work made use of the following software packages: \textsc{Astropy}~\cite{astropy2013,astropy2018,astropy2022}, \textsc{h5py}, \textsc{Jupyter}, \textsc{Matplotlib}~\cite{hunter2007}, \textsc{NumPy}~\cite{harris2020}, \textsc{pandas}~\cite{mckinney2010} and \textsc{SciPy}~\cite{virtanen2020}, in addition to the \gls{lisa} simulation tools cited in the main text.
% AI statement
\glspl{llm} were used in this work to provide feedback on and editing assistance with the manuscript.
They were also used to generate code snippets and to debug errors in the analysis scripts, which were based on code written by the authors.
All content was reviewed, verified and edited by the authors, who take full responsibility for this work.

\end{acknowledgments}

%--------------------------------------
% Appendix
%--------------------------------------
\appendix

% AET
%--------------------------------------
\section{Orthogonal coronagraphic TDI} \label{sec:AET}

% Motivation
The orthogonal basis ${(\mathcal{A}, \mathcal{E}, \mathcal{T})}$, first introduced in~\cite{PhysRevD.66.122002}, is designed to be an \textit{optimal basis} for \gls{lisa}.
Its Michelson analogue ${(A, E, T)}$ is commonly used in data analysis.
This is because, under simplifying assumptions~\cite{PhysRevD.66.122002}, the covariance matrix ${S(f)}$ is diagonal in this basis.
This is an interesting property because the inverse of the covariance matrix enters the Whittle likelihood used in data analysis.
If the covariance matrix is diagonal, the computation of the likelihood is greatly simplified.
It should be noted, however, that this diagonal property only holds under strong and unrealistic assumptions~\cite{PhysRevD.107.123531}.
Still, because the orthogonal basis ${(\mathcal{A}, \mathcal{E}, \mathcal{T})}$ relates to the Sagnac basis through a linear transformation, we can easily derive an orthogonal \gls{tdi} variable $\tilde{\kappa}^{\mathrm{orth}}$, as we show next. 

% Derivation
The orthogonal variables ${(\mathcal{A}, \mathcal{E}, \mathcal{T})}$ are linear combinations of the Sagnac variables, such that
\begin{equation}
    \mathcal{A} = \frac{\tilde{\gamma} - \tilde{\alpha}}{\sqrt{2}} \quad \mathcal{E} = \frac{\tilde{\alpha} - 2\tilde{\beta} + \tilde{\gamma}}{\sqrt{6}} \quad \mathcal{T} = \frac{\tilde{\alpha} + \tilde{\beta} + \tilde{\gamma}}{\sqrt{3}}.
\end{equation}
For the purpose of this derivation, we can alternatively use a transition matrix $P$ to express the change of basis from ${(\tilde{\alpha}, \tilde{\beta}, \tilde{\gamma})}$ to ${(\mathcal{A}, \mathcal{E}, \mathcal{T})}$ such that
\begin{equation}
    \begin{pmatrix}
        \mathcal{A} \\
        \mathcal{E} \\
        \mathcal{T}
    \end{pmatrix} = \underbrace{
    \begin{pmatrix}
        -1/\sqrt{2} & 0 & 1/\sqrt{2} \\
        1/\sqrt{6} & -2/\sqrt{6} & 1/\sqrt{6} \\
        1/\sqrt{3} & 1/\sqrt{3} & 1/\sqrt{3}
    \end{pmatrix}}_{P}
    \begin{pmatrix}
        \tilde{\alpha} \\
        \tilde{\beta} \\
        \tilde{\gamma}
    \end{pmatrix}.
\end{equation}
It follows that \cref{eq:kappa} can be rewritten as
\begin{equation}
    \tilde{\kappa} = \vec{A}(f, \delta, \alpha) \cdot \begin{pmatrix}
        \tilde{\alpha} \\
        \tilde{\beta} \\
        \tilde{\gamma}
    \end{pmatrix} = \underbrace{\vec{A}(f, \delta, \alpha) \cdot P^{-1}}_{\vec{A}^{\mathrm{orth}}(f,\delta,\alpha)} \begin{pmatrix}
        \mathcal{A} \\
        \mathcal{E} \\
        \mathcal{T}
    \end{pmatrix}.
\end{equation}
By setting ${\vec{A}^{\mathrm{orth}}(f,\delta,\alpha) = \vec{A}(f, \delta, \alpha) \cdot P^{-1}}$, we can define a $\tilde{\kappa}^{\mathrm{orth}}$ variable which takes as input the orthogonal Sagnac variables
\begin{equation}
    \vec{D}^{\mathrm{orth}}(f) = (\mathcal{A}(f), \mathcal{E}(f), \mathcal{T}(f))^T,
\end{equation}
 and reads
\begin{equation}
    \tilde{\kappa}^{\mathrm{orth}}(f, \delta, \alpha) = \vec{A}^{\mathrm{orth}}(f, \delta, \alpha) \cdot \vec{D}^{\mathrm{orth}}(f).
\end{equation}

% Comment
This orthogonal coronagraphic \gls{tdi} variable has been tested and implemented in the \textsc{NullStream} package.
It has the same coronagraphic property as $\tilde{\kappa}$.
The benefits and limitations of using $\tilde{\kappa}^{\mathrm{orth}}$ over $\tilde{\kappa}$ have not been studied in detail.
However, we could make similar arguments to those used to advocate for ${(\mathcal{A}, \mathcal{E}, \mathcal{T})}$ instead of ${(\tilde{\alpha}, \tilde{\beta}, \tilde{\gamma})}$ since the covariance matrix $S(f)$ enters the computation of ${S_{\kappa\kappa}(f, \delta, \alpha)}$ in \cref{eq:noise}.

% Params
%--------------------------------------
\section{Simulation parameters} \label{sec:params}

% Description
In this appendix, we provide the parameters of the sources used in the multiple-source analysis presented in Sec.~\ref{sec:superposition-multiple}.
This noiseless simulation contains one \gls{mbhb} from the \gls{ldc} Sangria dataset~\cite{sangria}, namely \gls{mbhb}~10, together with the 49 \glspl{vgb} of~\cite{kupfer2023lisa}, and is used to test whether the conclusions drawn from pairs of \glspl{gb} in Sec.~\ref{sec:superposition} hold in a more complex scenario.
The parameters of the \gls{mbhb} are given in \cref{tab:sim-params-mbhb}, and those of the \glspl{vgb} in \cref{tab:galactic-binaries}.

% Table: MBHB params
\begin{table}[tbp]
\begin{ruledtabular}
\begin{tabular}{c c c}
Parameter & \gls{mbhb}~10 & Units \\
\colrule
$M_1$ & $1.35 \times 10^{6}$ & $M_\odot$ \\
$M_2$ & $1.20 \times 10^{6}$ & $M_\odot$  \\
$M$ & $2.55 \times 10^{6}$ & $M_\odot$ \\
$z$ & $3.59$ & - \\
$\delta_\star$ & $1.29$ & rad \\
$\alpha_\star$ & $2.11$ & rad \\
\end{tabular}
\end{ruledtabular}
\caption{\label{tab:sim-params-mbhb}Parameters of the \gls{mbhb} used in the analysis presented in Sec.~\ref{sec:superposition-multiple} (see \cref{fig:mbhb-vgbs-delta}), based on~\cite{sangria}.
The data are simulated at a sampling rate of $0.2$\,Hz for 4 days ($69\,120$ data points).}
\end{table}

\begin{table*}[tbp]
\begin{ruledtabular}
\begin{tabular}{c c c c c c c}
Source & $\mathcal{A}$ & $f_0$ [Hz] & $\dot{f}$ [Hz/s] & $\delta_\star$ [rad] & $\alpha_\star$ [rad] & $\iota$ [rad] \\
\hline
HMCnc & ${4.25 \times 10^{-23}}$ & ${6.22 \times 10^{-3}}$ & ${7.49 \times 10^{-16}}$ & $-0.08$ & $2.10$ & $0.66$\\
ZTFJ1539 & ${9.40 \times 10^{-23}}$ & ${4.82 \times 10^{-3}}$ & ${2.54 \times 10^{-16}}$ & $1.15$ & $3.58$ & $1.47$\\
ZTFJ2243 & ${1.22 \times 10^{-22}}$ & ${3.79 \times 10^{-3}}$ & ${1.14 \times 10^{-16}}$ & $0.94$ & $0.23$ & $1.43$\\
V407Vul & ${9.38 \times 10^{-23}}$ & ${3.51 \times 10^{-3}}$ & ${8.28 \times 10^{-17}}$ & $0.82$ & $5.15$ & $1.05$\\
ESCet & ${9.52 \times 10^{-23}}$ & ${3.22 \times 10^{-3}}$ & ${5.54 \times 10^{-17}}$ & $-0.35$ & $0.43$ & $1.05$\\
SDSSJ0651 & ${1.58 \times 10^{-22}}$ & ${2.61 \times 10^{-3}}$ & ${2.63 \times 10^{-17}}$ & $0.10$ & $1.77$ & $1.52$\\
ZTFJ0538 & ${1.63 \times 10^{-22}}$ & ${2.31 \times 10^{-3}}$ & ${1.95 \times 10^{-17}}$ & $-0.06$ & $1.48$ & $1.49$\\
SDSSJ1351 & ${5.33 \times 10^{-23}}$ & ${2.13 \times 10^{-3}}$ & ${7.68 \times 10^{-18}}$ & $0.08$ & $3.64$ & $1.05$\\
AMCVn & ${2.80 \times 10^{-22}}$ & ${1.94 \times 10^{-3}}$ & ${6.06 \times 10^{-18}}$ & $0.65$ & $2.97$ & $0.75$\\
ZTFJ1905 & ${9.92 \times 10^{-23}}$ & ${1.94 \times 10^{-3}}$ & ${4.91 \times 10^{-18}}$ & $0.94$ & $5.13$ & $1.22$\\
SDSSJ1908 & ${6.48 \times 10^{-23}}$ & ${1.84 \times 10^{-3}}$ & ${3.86 \times 10^{-18}}$ & $1.07$ & $5.20$ & $0.26$\\
HPLib & ${1.55 \times 10^{-22}}$ & ${1.81 \times 10^{-3}}$ & ${2.52 \times 10^{-18}}$ & $0.09$ & $4.10$ & $0.52$\\
SDSSJ0935 & ${4.88 \times 10^{-22}}$ & ${1.68 \times 10^{-3}}$ & ${8.98 \times 10^{-18}}$ & $0.49$ & $2.29$ & $1.05$\\
J1239m2041 & ${1.59 \times 10^{-22}}$ & ${1.48 \times 10^{-3}}$ & ${4.89 \times 10^{-18}}$ & $-0.26$ & $3.45$ & $1.24$\\
CRBoo & ${1.23 \times 10^{-22}}$ & ${1.36 \times 10^{-3}}$ & ${1.06 \times 10^{-18}}$ & $0.31$ & $3.53$ & $0.52$\\
SDSSJ0634 & ${1.73 \times 10^{-22}}$ & ${1.26 \times 10^{-3}}$ & ${1.45 \times 10^{-18}}$ & $0.26$ & $1.69$ & $0.65$\\
V803Cen & ${1.93 \times 10^{-22}}$ & ${1.25 \times 10^{-3}}$ & ${1.06 \times 10^{-18}}$ & $-0.53$ & $3.77$ & $0.24$\\
4U1820-30 & ${1.36 \times 10^{-23}}$ & ${2.92 \times 10^{-3}}$ & ${2.50 \times 10^{-17}}$ & $-0.12$ & $4.80$ & $1.05$\\
SDSSJ2322 & ${7.03 \times 10^{-23}}$ & ${1.66 \times 10^{-3}}$ & ${2.72 \times 10^{-18}}$ & $0.15$ & $6.17$ & $1.05$\\
PTFJ0533 & ${8.14 \times 10^{-23}}$ & ${1.62 \times 10^{-3}}$ & ${3.96 \times 10^{-18}}$ & $-0.37$ & $1.45$ & $1.27$\\
ZTFJ2029 & ${8.35 \times 10^{-23}}$ & ${1.60 \times 10^{-3}}$ & ${3.63 \times 10^{-18}}$ & $0.58$ & $5.49$ & $1.51$\\
PTF1919 & ${3.14 \times 10^{-23}}$ & ${1.48 \times 10^{-3}}$ & ${1.37 \times 10^{-18}}$ & $1.20$ & $5.39$ & $1.05$\\
CX1751 & ${3.64 \times 10^{-23}}$ & ${1.46 \times 10^{-3}}$ & ${1.23 \times 10^{-18}}$ & $-0.11$ & $4.68$ & $1.05$\\
ZTFJ0722 & ${7.18 \times 10^{-23}}$ & ${1.41 \times 10^{-3}}$ & ${2.84 \times 10^{-18}}$ & $-0.70$ & $2.02$ & $1.56$\\
KLDra & ${3.59 \times 10^{-23}}$ & ${1.33 \times 10^{-3}}$ & ${7.68 \times 10^{-19}}$ & $1.37$ & $5.83$ & $1.05$\\
PTF0719 & ${3.66 \times 10^{-23}}$ & ${1.25 \times 10^{-3}}$ & ${5.79 \times 10^{-19}}$ & $0.46$ & $1.82$ & $1.05$\\
CPEri & ${3.62 \times 10^{-23}}$ & ${1.15 \times 10^{-3}}$ & ${4.00 \times 10^{-19}}$ & $-0.46$ & $0.74$ & $1.05$\\
SMSSJ0338 & ${1.21 \times 10^{-22}}$ & ${1.09 \times 10^{-3}}$ & ${8.17 \times 10^{-19}}$ & $1.04$ & $1.40$ & $1.20$\\
J2322p2103 & ${1.01 \times 10^{-22}}$ & ${1.04 \times 10^{-3}}$ & ${1.27 \times 10^{-18}}$ & $0.40$ & $0.00$ & $1.05$\\
SDSSJ0106 & ${7.63 \times 10^{-23}}$ & ${8.53 \times 10^{-4}}$ & ${3.80 \times 10^{-19}}$ & $-0.28$ & $0.20$ & $1.17$\\
SDSSJ1630 & ${1.38 \times 10^{-22}}$ & ${8.38 \times 10^{-4}}$ & ${6.72 \times 10^{-19}}$ & $1.10$ & $4.05$ & $1.05$\\
J1526m2711 & ${1.35 \times 10^{-22}}$ & ${8.27 \times 10^{-4}}$ & ${4.67 \times 10^{-19}}$ & $-0.14$ & $4.12$ & $1.05$\\
J1506m1125 & ${1.05 \times 10^{-22}}$ & ${7.16 \times 10^{-4}}$ & ${1.56 \times 10^{-19}}$ & $0.10$ & $3.97$ & $1.05$\\
SDSSJ1235 & ${1.13 \times 10^{-22}}$ & ${6.73 \times 10^{-4}}$ & ${1.51 \times 10^{-19}}$ & $0.31$ & $3.17$ & $0.47$\\
SDSSJ0923 & ${2.74 \times 10^{-22}}$ & ${5.15 \times 10^{-4}}$ & ${1.05 \times 10^{-19}}$ & $0.25$ & $2.33$ & $1.05$\\
CDm3011223 & ${3.94 \times 10^{-22}}$ & ${4.73 \times 10^{-4}}$ & ${1.44 \times 10^{-19}}$ & $-0.29$ & $3.86$ & $1.45$\\
SDSSJ1337 & ${4.39 \times 10^{-22}}$ & ${3.37 \times 10^{-4}}$ & ${1.86 \times 10^{-20}}$ & $0.80$ & $3.19$ & $0.23$\\
HD265435 & ${3.36 \times 10^{-22}}$ & ${3.36 \times 10^{-4}}$ & ${5.76 \times 10^{-20}}$ & $0.18$ & $1.77$ & $1.12$\\
ZTFJ1749 & ${4.33 \times 10^{-23}}$ & ${1.26 \times 10^{-3}}$ & ${1.73 \times 10^{-18}}$ & $0.57$ & $4.66$ & $1.49$\\
ZTFJ1946 & ${2.80 \times 10^{-23}}$ & ${9.93 \times 10^{-4}}$ & ${5.66 \times 10^{-19}}$ & $0.91$ & $5.38$ & $1.35$\\
ZTFJ0640 & ${4.96 \times 10^{-23}}$ & ${8.94 \times 10^{-4}}$ & ${5.46 \times 10^{-19}}$ & $-0.10$ & $1.74$ & $1.14$\\
ZTFJ2130 & ${7.80 \times 10^{-23}}$ & ${8.47 \times 10^{-4}}$ & ${6.05 \times 10^{-19}}$ & $0.95$ & $6.08$ & $1.51$\\
SDSSJ0822 & ${6.84 \times 10^{-23}}$ & ${8.23 \times 10^{-4}}$ & ${4.81 \times 10^{-19}}$ & $0.19$ & $2.11$ & $1.53$\\
ZTFJ1901 & ${8.29 \times 10^{-23}}$ & ${8.21 \times 10^{-4}}$ & ${4.07 \times 10^{-19}}$ & $1.30$ & $5.35$ & $1.52$\\
SDSSJ1043 & ${2.45 \times 10^{-23}}$ & ${7.30 \times 10^{-4}}$ & ${2.60 \times 10^{-19}}$ & $-0.04$ & $2.80$ & $1.05$\\
ZTFJ2320 & ${4.12 \times 10^{-23}}$ & ${6.03 \times 10^{-4}}$ & ${1.30 \times 10^{-19}}$ & $0.66$ & $0.15$ & $1.47$\\
WD0957 & ${2.55 \times 10^{-22}}$ & ${3.80 \times 10^{-4}}$ & ${2.23 \times 10^{-20}}$ & $-1.17$ & $3.64$ & $1.31$\\
J0221p1710 & ${1.84 \times 10^{-22}}$ & ${3.78 \times 10^{-4}}$ & ${2.70 \times 10^{-20}}$ & $0.05$ & $0.67$ & $1.55$\\
J2013m1310 & ${1.33 \times 10^{-22}}$ & ${3.76 \times 10^{-4}}$ & ${3.15 \times 10^{-20}}$ & $0.11$ & $5.28$ & $1.05$\\
\end{tabular}
\end{ruledtabular}
\caption{\label{tab:galactic-binaries}Parameters of the \glspl{vgb}~\cite{kupfer2023lisa} used in the analysis presented in Sec.~\ref{sec:superposition-multiple} (see \cref{fig:mbhb-vgbs-delta}), based on~\cite{kupfer2023lisa}.
The polarization $\Psi$ and initial phase $\phi$ are set to 0 for all sources.
The parameters ${\{\Psi, \delta_\star, \alpha_\star\}}$ are given in the \gls{icrs} frame, following \textsc{LISA GW Response}'s convention.
The data are simulated at a sampling rate of $0.2$\,Hz for 4 days ($69\,120$ data points).}
\end{table*}

% Calibration
%--------------------------------------
\section{Calibration} \label{sec:calibration}

% Intuition
Before discussing the calibration of $z$, we clarify why ${z(\delta, \alpha) \sim \chi^2_{2N_f}}$, as stated in the main text.
This is not a rigorous proof (see~\cite{PhysRevD.63.042003}), but rather an intuitive argument.
In its simplest form, a $\chi^2$ statistic is defined as~\cite{Cowan}
\begin{equation}
    \sum_{i} \frac{(x_i-\mu_i)^2}{\sigma_i^2} \sim \chi^2_N, \label{eq:chi2}
\end{equation}
for $\{x_i\}$ independent measurements following ${x_i \sim \mathcal{N}(\mu_i,\sigma_i^2)}$ for integer ${i \in [0,N-1]}$.
In reality, under $H_0$, $z(\delta, \alpha)$ takes the same form.
Indeed, assuming that the noise in the data is stationary and Gaussian with zero mean, we find that 
\begin{equation}
    \sigma_i^2(\delta, \alpha) = \langle |\tilde{\kappa}(f_i, \delta, \alpha)|^2 \times dt^2 \rangle = \frac{S_{\kappa\kappa}(f_i,\delta,\alpha)}{2 df}.
\end{equation}
This is just the standard result used in the Whittle likelihood (see for instance~\cite{burke_extreme_2021}) applied to coronagraphic \gls{tdi}, which is a linear combination of standard \gls{tdi} variables.
Also note that $S_{\kappa\kappa}$, given in \cref{eq:noise}, includes the correlations between channels ${(\tilde{\alpha}, \tilde{\beta}, \tilde{\gamma})}$.
Then we recognize in
\begin{equation}
    z(\delta,\alpha) = \sum_i 4 \frac{|\tilde{\kappa}(f_i, \delta, \alpha)|^2 \times dt^2}{S_{\kappa\kappa}(f_i, \delta, \alpha)} df, \label{eq:z-app}
\end{equation}
the variance $\sigma_i^2$ in the denominator and the data with zero mean in the nominator, with an additional factor of 2, following the same structure as \cref{eq:chi2}.
In addition, taking the mean of \cref{eq:z-app}, we obtain
\begin{equation}
    \langle z(\delta,\alpha) \rangle = \sum_i 4 \frac{\langle|\tilde{\kappa}(f_i, \delta, \alpha)|^2 \times dt^2\rangle}{S_{\kappa\kappa}(f_i, \delta, \alpha)} df = 2 N_f,
\end{equation}
as expected for a $\chi^2_{2N_f}$ distribution.
For a noise only dataset, any significant deviation from $2N_f$ indicates a calibration problem.

% Context
To support this argument with empirical evidence, we show the analysis of the 1\,000 noise simulations mentioned in Sec.~\ref{sec:analysis}.
These simulations were generated with parameters given in \cref{tab:noise-sims} under simulation D.
For each realization, we compute the statistic $z(\delta, \alpha)$ at an arbitrary position ${(\delta=1.0, \alpha=0.5)}$\,rad.
This allows us to obtain the distribution of $z(\delta, \alpha)$, which we expect to be $\chi^2_{2N_f}$ distributed.
Any deviation from the expected distribution for a controlled noise-only simulation indicates a calibration error in the computation of $z$.
It is known that calibration problems may arise when analyzing real data with null-stream channels and have been discussed in the context of ground-based detectors in~\cite{ajith_null-stream_2006}.
For simulated data, this serves primarily as a sanity check.

% Figure: histograms
\begin{figure}[tbp]
    \centering
    % Noise
    \begin{subfigure}[t]{\columnwidth}
        \centering
        \includegraphics[width=0.96\linewidth]{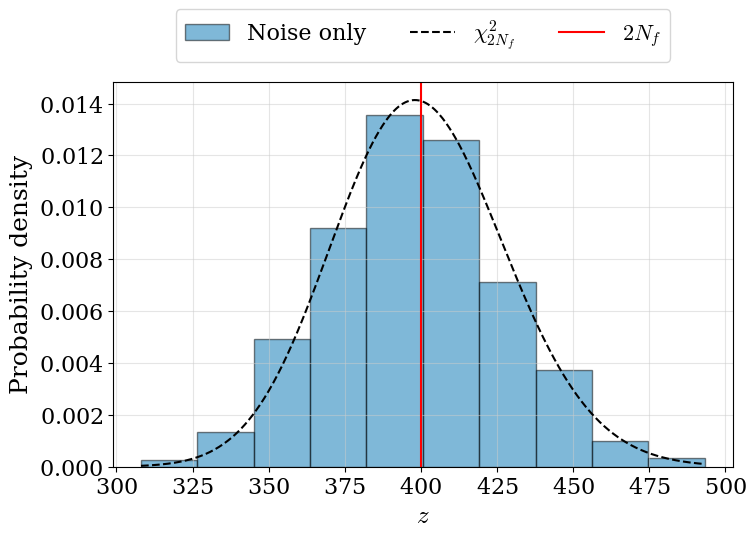}
        \caption{\label{fig:chi2-noise}Theoretical and empirical $z$ distribution obtained from 1\,000 noise-only simulations (see simulation D in \cref{tab:noise-sims}) at position ${(\delta=1.0, \alpha=0.5)}$\,rad. The $z$ statistic was computed over the interval ${[2.77, 3.23]}$\,mHz centered around ${3.00}$\,mHz. The histogram has mean 398 and standard deviation 29.}
    \end{subfigure}
    \vspace{0.3cm}
    % On source
    \begin{subfigure}[t]{\columnwidth}
        \centering
        \includegraphics[width=0.96\linewidth]{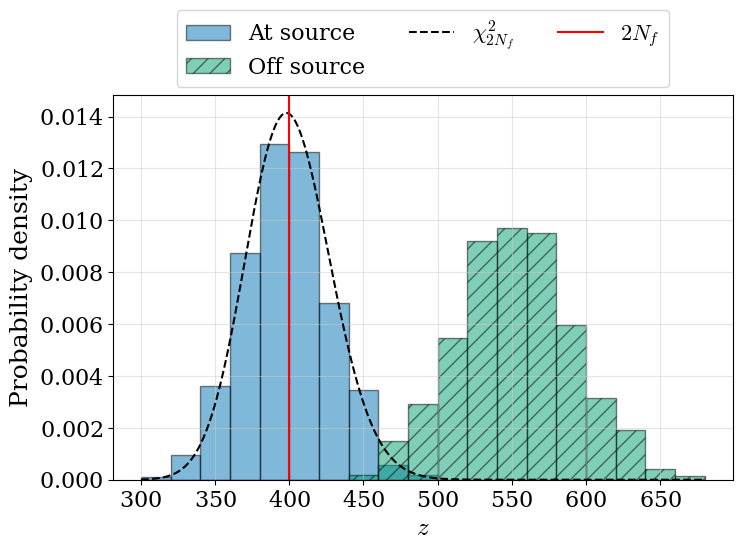}
        \caption{\label{fig:chi2-source}Theoretical and empirical $z$ distribution obtained from 1\,000 \gls{gb} simulations with noise (see simulation D in \cref{tab:noise-sims} for noise parameters and \cref{tab:sim-params-gbs} for source parameters). The \gls{gb} used in this case has $\mathcal{A} = 8.56 \times 10^{-20}$, corresponding to \gls{snr} $8\,701$. The plain blue histogram shows the distribution of $z$ at the position of the source ${(\delta_\star=0.92, \alpha_\star=5.95)}$\,rad in the \gls{icrs} frame, i.e., at source. The hatched green histogram shows the analog distribution at position ${(\delta=0.99,\alpha=0.53)}$\,rad, i.e., off source. The $z$ statistic was computed over the interval ${[3.56, 4.02]}$\,mHz centered around ${f_0=3.79}$\,mHz. The off source histogram has mean 552 and standard deviation 39, while the at source histogram has mean 398 and standard deviation 29.}
    \end{subfigure}

    \caption{\label{fig:chi2}Theoretical and empirical $z$ distributions obtained for two sets of simulations: noise only and a \gls{gb} injection with instrumental noise. In both cases $z$ was computed for 200 frequency bins with bin size ${df = 2.33\,\mu}$Hz.}
\end{figure}

% Noise only
In \cref{fig:chi2-noise}, we show the histogram obtained from the noise-only dataset.
We find a good agreement between the histogram and the theoretical $\chi^2_{2N_f}$ distribution.
For this distribution, we expect a mean of ${2N_f = 400}$ and a standard deviation of ${\sigma = \sqrt{2 \times 2N_f} \approx 28}$.
We obtained a mean of 398 and a standard deviation of 29 in our analysis (see \cref{fig:chi2-noise}).
This confirms that the method is well calibrated.

% Noise + GB
For comparison, in a second dataset, we add a \gls{gw} injection to each of the 1\,000 noise simulations of the first dataset.
The injection is a synthetic \gls{gb} whose parameters are given in \cref{tab:sim-params-gbs}.
For this example, we chose an injection with high amplitude of ${\mathcal{A} = 8.56 \times 10^{-20}}$, which corresponds to \gls{snr} $8\,701$ (see \cref{tab:violin-stats}), and sky position $(\delta_\star=0.92, \alpha_\star=5.95)$\,rad.
Unlike in the noise-only case, the empirical distribution at ${(\delta=0.5, \alpha=1.0)}$\,rad is far from a $\chi^2_{2N_f}$ distribution, as shown in \cref{fig:chi2-source}.
For reference, the empirical off source distribution has mean 552, ${5.4 \sigma}$ above ${2N_f = 400}$.

\begin{figure}[tbp]
    \centering
    \includegraphics[width=\linewidth]{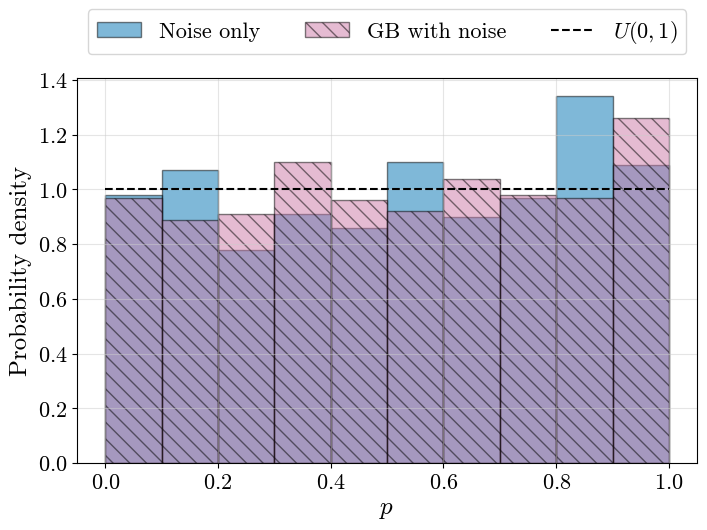}
    \caption{\label{fig:pdist}Theoretical and empirical distributions of the $p$ value obtained from the $z$ values shown in \cref{fig:chi2}.
    We consider two datasets: $1\,000$ \textit{noise-only} simulations and $1\,000$ simulations of a \textit{\gls{gb} with noise} (see simulation D in \cref{tab:noise-sims} for noise parameters and \cref{tab:sim-params-gbs} for source parameters).
    The plain blue histogram shows the distribution of $p$ for the noise-only simulations at arbitrary position ${(\delta=1.0,\alpha=0.5)}$ rad.
    The hatched pink histogram shows the distribution of $p$ at the position of the source ${(\delta_\star=0.92,\alpha_\star 
    =5.95)}$\,rad in the \gls{icrs} frame.
    The \gls{gb} used in this case has ${\mathcal{A} = 8.56 \times 10^{-20}}$, corresponding to \gls{snr} $8\,701$.
    The dashed line is the expected uniform distribution $U(0,1)$.}
\end{figure}

% p-value distribution
Going one step further in the analysis described in Sec.~\ref{sec:method}, we compute the $p$ values associated to the $z$ values from \cref{fig:chi2}.
As mentioned in the main text, we expect $p$ values to be uniformly distributed under the noise-only hypothesis $H_0$~\cite{Cowan}.
We show that this indeed the case in the simulations analyzed in Sec.~\ref{sec:analysis}.
In \cref{fig:pdist}, we show two histograms which represent the $p$ values obtained from the two datasets already described.
For the noise-only dataset, the histogram is taken at an arbitrary position ${(\delta=1.0, \alpha=0.5)}$\,rad, while for the dataset containing an injection the histogram is taken at the position of the source ${(\delta_\star=0.92, \alpha_\star=5.95)}$\,rad.
In both cases, we recover an uniform distribution which indicates that, at the position of the source, the data behave as noise, meaning that the \gls{gw} signal has been successfully suppressed.
Precisely because $p$ is uniformly distributed at the position of the source, it can take any value between $0$ and $1$ with equal probability and as a consequence fail the threshold test ${p(\delta_\star, \alpha_\star) > \beta}$ applied in Sec.~\ref{sec:method} despite a successful suppression of the \gls{gw} signal in the data.

% Figure: p distribution

%--------------------------------------
% References
%--------------------------------------
\bibliography{references}

\end{document}